\documentclass[lettersize,journal]{IEEEtran}
\usepackage{amsmath,amsfonts}
\usepackage{algorithmic}
\usepackage{array}
\usepackage[caption=false,font=normalsize,labelfont=sf,textfont=sf]{subfig}
\usepackage{textcomp}
\usepackage{stfloats}
\usepackage{url}
\usepackage{verbatim}
\usepackage{graphicx}
\usepackage{forest}
\usepackage{xcolor}
\usepackage{hyperref}
\usepackage{booktabs}
\usepackage{threeparttable}
\def\BibTeX{{\rm B\kern-.05em{\sc i\kern-.025em b}\kern-.08em
    T\kern-.1667em\lower.7ex\hbox{E}\kern-.125emX}}
\usepackage{balance}

\usepackage{booktabs}
\usepackage{multirow}

\forestset{
    orange/.style={fill=orange!20, draw=orange!60!black, text color=orange!80!black},
    green/.style={fill=green!20, draw=green!60!black, text color=green!60!black},
    purple/.style={fill=purple!20, draw=purple!60!black, text color=purple!80!black},
    beige/.style={fill=yellow!15, draw=yellow!60!black},
}

\begin{document}
\title{A Survey of Large Language Models for Perception and Measurement of Human Psychology}
\author{\textbf{Yudong~Li}\textsuperscript{1}, \textbf{Xiaoyi~Chen}\textsuperscript{1}, \textbf{Jiawei~Cai}\textsuperscript{1}, \textbf{Zehao~Zhong}\textsuperscript{1}, \textbf{Haoyang~Yang}\textsuperscript{1}, \textbf{Huajin~Tang}\textsuperscript{4}, and \textbf{Linlin~Shen}\textsuperscript{1,2,3*}\\
\textsuperscript{1}School of Artificial Intelligence, Shenzhen University, China\\
\textsuperscript{2}Department of Computer Science, University of Nottingham Ningbo China, Zhejiang, China\\
\textsuperscript{3}Guangdong Provincial Key Laboratory of Intelligent Information Processing, \\ Shenzhen University, China \\%
\textsuperscript{4}College of Computer Science and Technology, The State Key Lab of Brain-Machine Intelligence, \\Zhejiang University, China%

\thanks{*Corresponding author: Linlin Shen (e-mail: llshen@szu.edu.cn).}}

\markboth{IEEE Transactions on Cognitive and Developmental Systems}%
{How to Use the IEEEtran \LaTeX \ Templates}

\maketitle

\begin{abstract}
Against the backdrop of the rapid advancement of Large Language Models (LLMs), their application in the field of psychology has garnered significant academic attention. A central issue is whether LLMs possess the capability to accurately perceive and measure complex, latent human psychological constructs, such as personality, emotions, and cognitive states. This paper provides a systematic review focused on the use of LLMs as instruments for human psychological measurement. To organize this domain, we propose a comprehensive analytical framework structured around three critical dimensions: Theoretical Plausibility (why measurement might be possible), Measurement Methodology (how to measure), and Application Effectiveness (what has been measured). We first explore the theoretical foundations supporting LLM-based measurement, examining the debate on their emergent cognitive properties from a psychometric perspective. Next, we systematically analyze existing measurement paradigms, categorizing them into active conversational assessment, passive natural language analysis, and multimodal fusion. Subsequently, we review the practical effectiveness and limitations of LLMs in core application areas, including personality trait assessment and mental health evaluation. Distinct from prior reviews focusing on general applications or the ``psychology'' of LLMs themselves, this paper centers on the psychometric properties of LLMs as measurement tools.
\end{abstract}


\section{Introduction}

\begin{figure*}[h]
    \centering
    \includegraphics[width=0.9\linewidth]{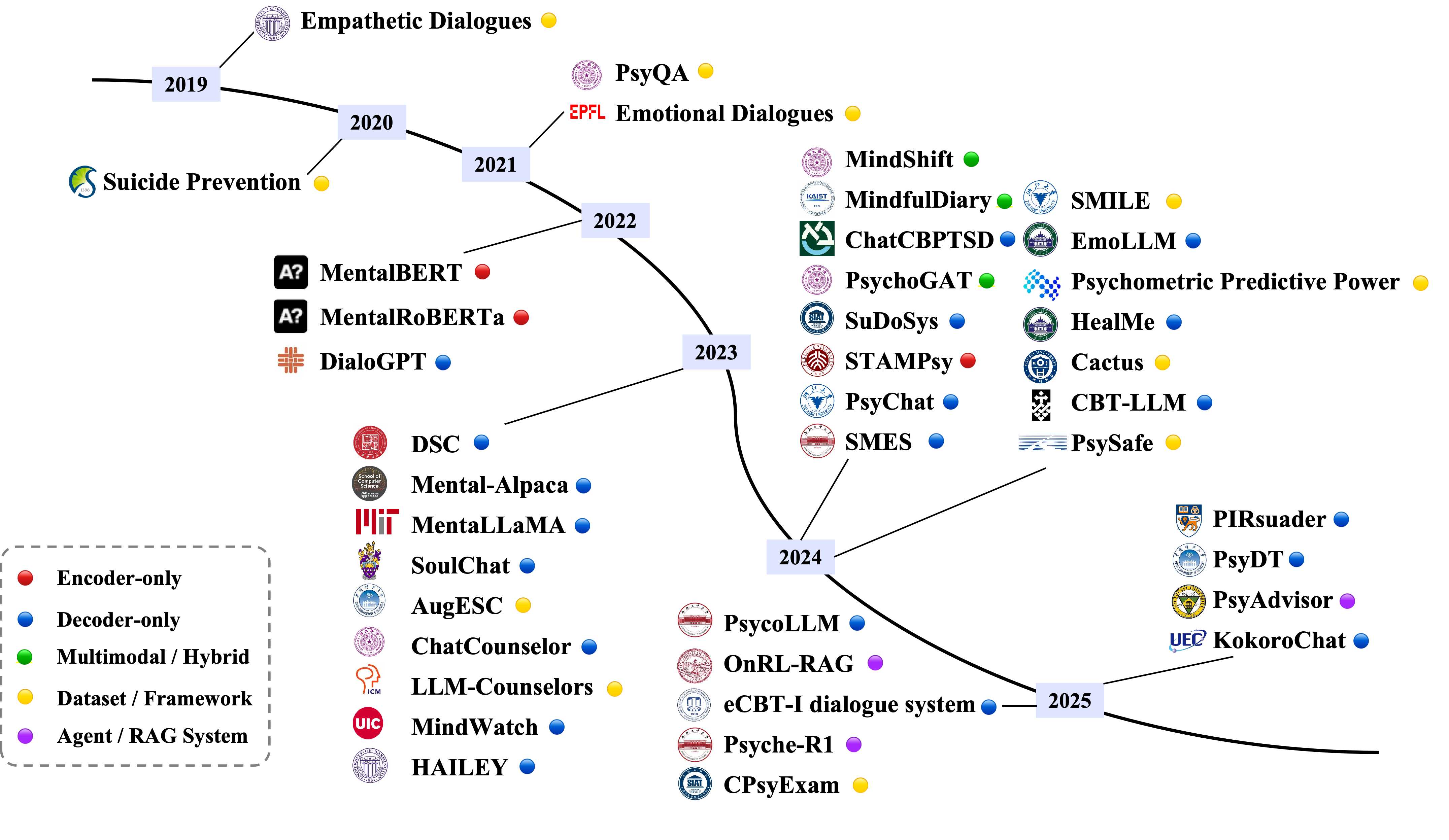}
    \caption{An overview of psychology domain-specific large language models form recent years.}
    \label{fig:timeline}
\end{figure*}

As Large Language Models (LLMs)~\cite{achiam2023gpt, google2025gemini2, guo2025deepseek} continue to advance, their applications have expanded into specialized domains, among which psychology has attracted growing attention. LLMs can process and generate human-like text, and their ability to perform tasks analogous to human cognition has shown potential in mental health support, cognitive assessment, and social dialogue. Domain-specialized models have demonstrated effectiveness in psychological counseling, emotion recognition, and behavioral prediction~\cite{chen2023soulchat, qiu2023smile, hu2024psycollm}. Figure~\ref{fig:timeline} illustrates the rapid growth of psychology-oriented LLMs in recent years.

However, these developments raise a central question: Can LLMs perceive and measure complex, latent human psychological attributes such as personality traits, emotional states, and cognitive styles? This survey addresses this question through three dimensions: \textit{theoretical foundations}, \textit{measurement methodology}, and \textit{application effectiveness}.

\textbf{Theoretical Plausibility} - \textit{why measurement might be possible}: A fundamental debate persists within the community as to whether LLMs possess cognitive properties that make psychological measurement meaningful. Some view LLMs as sophisticated statistical learners that generate language by exploiting correlations within large-scale corpora, without true comprehension or grounded understanding \cite{bender2021dangers}. From this perspective, their apparent performance in tasks involving reasoning, empathy, or social cognition might be better explained as emergent artifacts of statistical-level pattern recognition. However, growing evidence suggests that LLMs can exhibit behaviors resembling aspects of human cognition, such as theory of mind, emotion recognition, and social reasoning. Recent studies demonstrate that LLMs can perform well on classical false-belief tasks and other benchmarks traditionally used to evaluate human social cognition, indicating that they may encode or approximate latent psychological constructs \cite{kosinski2023theory, chen2024tombench, ullman2023large}. This highlights the need for rigorous theoretical grounding when positioning LLMs as tools for psychological assessment.

\textbf{Measurement Methodology} - \textit{how to measure}: Even if LLMs approximate psychological constructs, developing reliable and valid measurement frameworks presents substantial challenges. Recent studies have proposed methods that leverage zero-shot prompting, embeddings, and fine-tuned models to infer psychological states from text data, ranging from personality traits to emotional engagement. For instance, embedding-based approaches have been shown reliability and meaningful correlations with established linguistic markers \cite{maharjan2025psychometric}. Similarly, LLM-generated ratings of therapeutic interactions have achieved strong internal consistency and convergent validity with human coders \cite{wilf2024think}. For this topic, how to address biases in model outputs, how to ensure temporal stability across repeated measurements, and how to design benchmark datasets that reflect diverse cultural and contextual conditions still remain relatively unexplored. These challenges underscore the need for psychometrically grounded standards in LLM-based assessment.

\textbf{Application} - \textit{what has been measured}:
LLMs have been deployed in tasks such as emotion recognition, personality assessment, and cognitive evaluation. While some studies report human-level performance in specific domains  \cite{sabour2024emobench}, there are still significant limitations, such as reduced stability across time \cite{sileo2023mindgames}, inconsistent handling of reverse-coded questionnaire items \cite{salecha2024large}, and systematic biases toward socially desirable responses \cite{suhr2023challenging}. Applications in psychotherapy and mental health contexts further highlight the opportunity and risk: while LLMs can efficiently generate engagement scores from therapy transcripts with psychometric reliability, their deployment raises pressing ethical concerns, including privacy, fairness, and the potential consequences of misinterpretation in clinical settings \cite{bodrovza2024personality}. These studies suggest that while LLMs may complement traditional psychological tools under controlled conditions, they cannot yet be considered reliable substitutes for human-administered assessments in high-stakes contexts.

\begin{figure*}[!t]
    \centering
    \includegraphics[width=\linewidth]{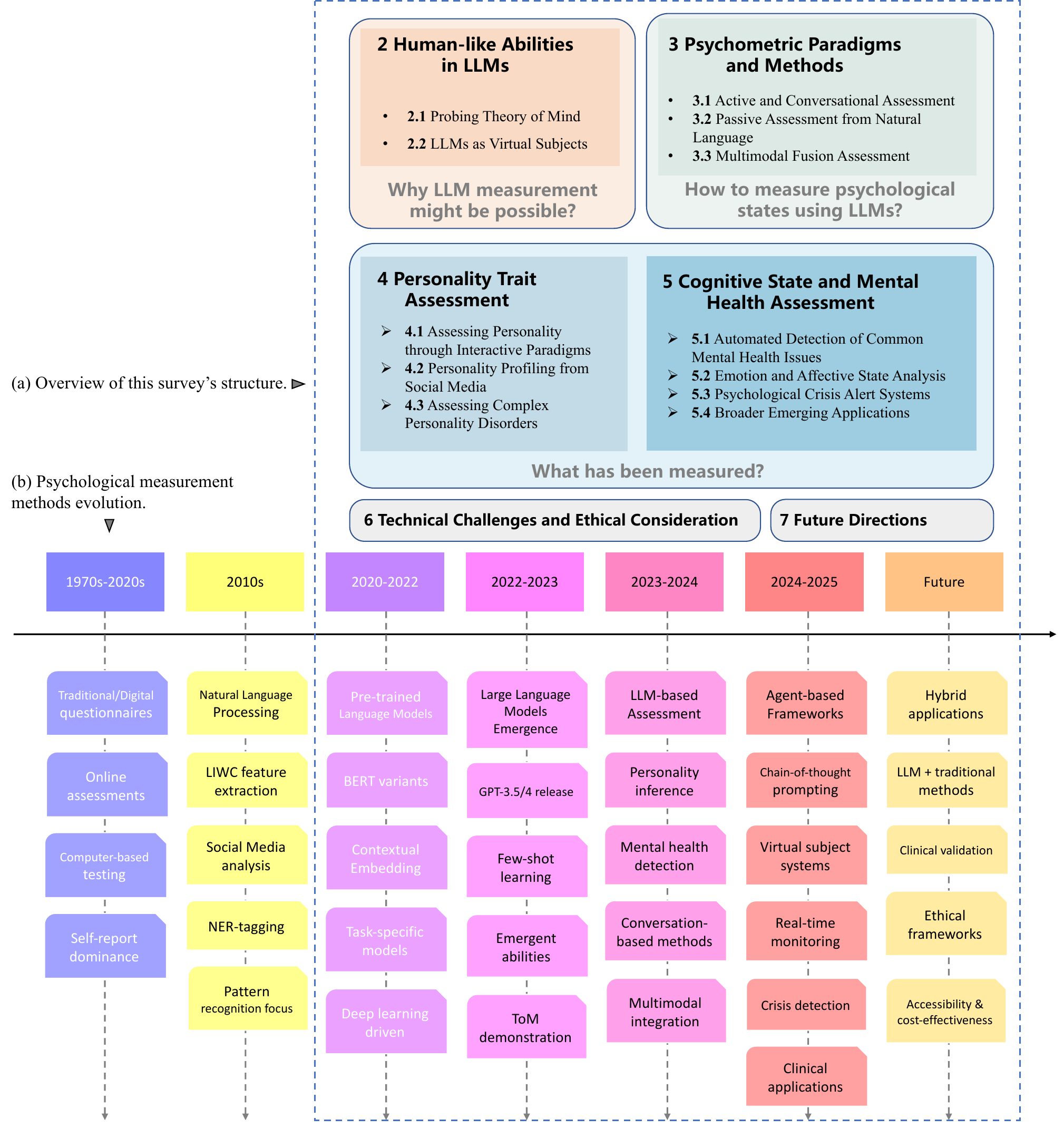}

    \caption{Overview of this survey. (a) Section structure. (b) Evolution of psychological measurement methods: from traditional questionnaires (pre-2020), through pre-trained language models such as BERT for text-based assessment (2020--2022), to LLM-based paradigms including GPT-3.5/4 for active, passive, and multimodal measurement (2023--present).}
    \label{fig:method}
\end{figure*}

This paper provides a systematic review of LLM-based perception and measurement of human psychological indicators. Recent surveys have examined related topics from different angles, including assessing LLMs' own psychological traits~\cite{dong2025humanizing, ye2025large} and surveying LLM applications in psychology broadly~\cite{ke2025exploring, guo2024large}. Our review adopts a distinct focus: we examine LLMs specifically as instruments for measuring human psychological states, analyzing the measurement pipeline through the lens of psychometric theory.

This paper is organized as follows: Section~\ref{ch:02} establishes theoretical foundations by examining emergent human-like cognitive abilities in LLMs. Section~\ref{ch:03} presents a systematic analysis of psychometric paradigms, categorizing approaches into active conversational assessment, passive natural language analysis, and multimodal fusion. Sections~\ref{ch:04} and~\ref{ch:05} review practical applications in personality trait assessment and mental health evaluation, respectively. Section~\ref{ch:06} addresses technical challenges and ethical considerations, including reliability, interpretability, privacy, and fairness. Finally, Section~\ref{ch:07} explores future directions in technical development, application domains, and standardization efforts.
\section{Human-like Abilities in LLMs}
\label{ch:02}

Before deploying LLMs as psychometric instruments, a fundamental question must be addressed: Do LLMs possess cognitive properties sufficient to support meaningful psychological measurement? The debate on whether LLMs are ``true understanders'' or ``advanced statistical imitators'' remains open~\cite{y2022large,gurnee2023language,mitchell2023debate,bender2020climbing,yan2024large}. Nevertheless, growing empirical evidence suggests that their capabilities can be examined at the functional and behavioral levels using methods from psychology. The central observation is that advanced LLMs have developed human-like abilities that closely approximate social cognitive processes, providing a theoretical basis for their use in psychological measurement. This section examines two complementary dimensions. Section~\ref{sec:tom} addresses outward understanding: the ability to infer others' mental states, assessed through Theory of Mind (ToM) tasks~\cite{van2023theory,kosinski2023theory,baron1997mindblindness}. Section~\ref{sec:virtual_subjects} examines inward simulation: the capacity to enact specific psychological roles as virtual subjects.

\subsection{Probing Theory of Mind}
\label{sec:tom}
This section examines the theoretical basis for using LLMs in psychological measurement: the spontaneous emergence of Theory of Mind (ToM) capabilities. ToM refers to the ability to infer others' mental states, including thoughts, feelings, and beliefs, and to predict their behavior accordingly~\cite{schlinger2009theory}. Long regarded as a core mechanism for social interaction~\cite{moghaddam2023boosting,sap2022neural,strachan2024testing,kosinski2024evaluatinglargelanguagemodels} and even a uniquely human trait~\cite{baron1997mindblindness}, ToM has recently been observed to emerge in LLMs without targeted training~\cite{kosinski2024evaluatinglargelanguagemodels}. This capability appears as a byproduct of scaling. Regardless of its origin, its functional presence provides a cognitive basis for psychological measurement applications. 

The validation of LLMs' ToM abilities draws on classic psychology experimental paradigms. Among these, the false belief task and the strange stories test are two foundational paradigms for assessing basic and advanced ToM abilities, while other tasks (e.g., faux pas recognition, higher-order ToM) serve as extensions that probe more nuanced social reasoning. Table~\ref{tab:tom_tasks} summarizes representative results across these paradigms.

\textbf{False Belief Task.} This paradigm assesses whether an individual understands that others may hold beliefs inconsistent with reality~\cite{wimmer1983beliefs}. Standard tests include the Sally-Anne Test~\cite{baron1985does} and the Smarties Task~\cite{perner1987three}. Kosinski~\cite{kosinski2023theory} found that GPT-3.5 (davinci-003) achieves 93\% accuracy on false belief tasks, comparable to the level of a nine-year-old child. Under more rigorous control conditions, GPT-4 performed at the level of 6--7-year-old children~\cite{kosinski2024evaluatinglargelanguagemodels}. Van Duijn et al.~\cite{van2023theory} confirmed that GPT-4, PaLM-2, and LLaMA surpass the performance of 7--10-year-old children on first-order false belief tasks, though accuracy decreases on second-order tasks.

\textbf{Strange Stories Test.} To probe more advanced ToM abilities, researchers have used the Strange Stories Test, which requires understanding nonliteral mental states such as sarcasm, metaphors, and white lies. Strachan et al.~\cite{strachan2024testing} found that GPT-4 performs comparably to adult humans on these tasks. Chen et al.~\cite{chen2024tombench} constructed ToMBench, a bilingual (English--Chinese) benchmark covering eight task categories including Strange Stories and Faux Pas, enabling systematic evaluation of LLM social intelligence.

Neuroscience research offers complementary evidence at the representational level. Goldstein et al.~\cite{goldstein2022shared} showed that LLM internal representations reliably predict brain neural activity patterns during language processing, suggesting convergent computational principles despite distinct biological substrates. Taken together, the behavioral and computational evidence indicates that LLMs have reached key developmental benchmarks in structured ToM tasks.

\begin{table}[ht]
\centering
\caption{Comparison of LLM Performance on ToM Tasks}
\label{tab:tom_tasks}
\renewcommand{\arraystretch}{1.2}
\begin{tabular}{@{}lll@{}}
\toprule
\textbf{Test} & \textbf{Model(s)} & \textbf{Performance} \\
\midrule
False Belief \cite{kosinski2023theory} & GPT-3.5 & 9-yr-old level \\
& GPT-4 & 6-7 yr-old level \\
False Belief \cite{van2023theory} & GPT-4, PaLM-2 & $>$7-10 yr-old children \\
Strange Stories \cite{strachan2024testing} & GPT-4 & $\approx$ or $>$ adult humans \\
ToM-bench \cite{chen2024tombench} & GPT-4 & 10.1\% below human (85.4\%) \\
Hi-ToM \cite{street2024llms} & Claude, Guanaco & adult-level performance \\
\bottomrule
\end{tabular}
\end{table}

However, the validity of existing ToM benchmarks has itself been questioned. Riemer et al.~\cite{riemer2025tom_broken} argued that most benchmarks only measure ``literal'' ToM (predicting behavior from stated beliefs) rather than ``functional'' ToM (adapting to partners in context). Their experiments showed that strong literal ToM performance does not guarantee functional ToM, and that many open-source LLMs fail under trivially altered task conditions. A recent ACL survey~\cite{chen2025tom_survey_acl} further emphasized the need for contamination-resistant benchmarks and multi-lingual evaluation. These findings suggest that current ToM results should be interpreted cautiously; structured tasks may overstate the robustness of LLMs' social reasoning.

In summary, ToM-like abilities in LLMs have been repeatedly demonstrated on standard benchmarks, but their depth and generalizability remain uncertain. The functional capability provides a reasonable starting point for exploring LLM-based psychological measurement, while acknowledging that performance on existing tests may not fully reflect genuine social understanding.

\subsection{LLMs as Virtual Subjects}
\label{sec:virtual_subjects}

The role-playing ability discussed above not only influences ToM performance but also introduces methodological complexity. A natural question follows: Can this simulation capability be harnessed as a research tool rather than treated as a confound? Recent work on ``LLM as a virtual subject'' provides a positive answer. This section introduces the core methodology, reviews key applications, and discusses the practical value and limitations of this approach.

The core methodology is persona prompting. Researchers construct a virtual subject by providing the LLM with a role description combining demographic variables~\cite{goldberg1998demographic} (age, gender, race, education) and psychological traits~\cite{allport1937personality} (personality profiles, values, mental states). Systematically varying these parameters yields a controllable virtual subject pool, which Argyle et al.~\cite{argyle2023out} termed ``silicon samples.''

At the practical application level, this approach has been applied across diverse domains. Argyle et al.~\cite{argyle2023out} showed that GPT-3 conditioned on ANES demographic backgrounds reproduced group-level political attitudes with high fidelity. Wang et al.~\cite{wang2025evaluating} deepened simulation granularity to individual personalities, finding that GPT-4 playing Big Five trait profiles even surpassed real human data on several psychometric indicators. The paradigm has since been extended to cross-cultural moral decision-making~\cite{tuna2024effects,kovavc2023large} and clinical virtual patient simulation~\cite{wang2024patient}.

The ``LLM as a virtual subject'' approach offers three practical benefits for psychological research. First, it reduces the cost and time of data collection, enabling rapid iteration of large-scale studies~\cite{argyle2023out,broska2025mixed}. Second, it improves controllability: researchers can precisely manipulate demographic and psychological variables while holding others constant, facilitating causal inference~\cite{wang2025evaluating,sorokovikova2024llms}. Third, it extends ethical boundaries by enabling the simulation of sensitive scenarios that would be difficult or impermissible to conduct with real participants~\cite{wang2024patient,cheung2025large}. These simulations are not substitutes for studies with real individuals; rather, they serve as tools for hypothesis generation and population-level exploration.

\begin{table*}[h]
\centering
\caption{Summary of Works on LLMs as Virtual Subjects}
\label{tab:virtual_subjects}
\renewcommand{\arraystretch}{1.2}
\begin{tabular}{p{0.8cm}p{3.2cm}p{4.5cm}p{6.8cm}}
\toprule
\textbf{Year} & \textbf{Study} & \textbf{Method} & \textbf{Construct Simulated} \\
\midrule
2023 & Argyle et al. \cite{argyle2023out} & Demographic persona prompting & Political attitudes and voting behavior (ANES survey) \\ 

2023 & Ramezani et al. \cite{ramezani2023knowledge} & Cultural context prompting & Cross-cultural moral norms across 40+ countries \\

2024 & Wang et al. \cite{wang2024patient} & Clinical persona prompting & Psychiatric symptoms for therapeutic training \\

2024 & Reichenpfader et al. \cite{reichenpfader2024simulating} & Hybrid demographic-clinical prompting & Diverse patient populations with medical conditions \\

2024 & Sorokovikova et al. \cite{sorokovikova2024llms} & Psychological trait prompting & Big Five personality traits \\

2024 & Moon et al. \cite{moon2024virtual} & Persona anthology engineering & Social interaction behaviors and responses \\

2025 & Wang et al. \cite{wang2025evaluating} & Psychological trait prompting & Big Five personality traits \\

2025 & Cheung et al. \cite{cheung2025large} & Moral dilemma prompting & Cognitive biases in ethical decision-making \\

2025 & Broska et al. \cite{broska2025mixed} & Mixed subjects design & Social attitudes with human-silicon sample integration \\

2025 & Toubia et al. \cite{toubia2025database} & Digital twin engineering & Individual behavioral patterns  \\

\bottomrule
\end{tabular}
\end{table*}

\subsection{Summary and Position Statement}

Three conceptually distinct research settings that appear in the literature and throughout this survey: (1)~\textit{using LLMs to assess human psychology}, where the model serves as a measurement instrument applied to human-generated data; (2)~\textit{using LLMs to simulate human psychology}, where the model acts as a virtual subject embodying specified psychological profiles; and (3)~\textit{assessing the psychological properties of LLMs themselves}, where the model is the object of measurement. These settings require different evaluation standards and should not be conflated. This survey focuses primarily on settings~(1) and~(2), while referencing setting~(3) only where it informs measurement validity.

Based on the evidence reviewed in this section, we offer the following position. Current LLMs demonstrate functional ToM abilities sufficient to support structured psychological measurement tasks such as trait inference and symptom screening. However, these abilities remain context-dependent and fragile under adversarial or out-of-distribution conditions. LLMs can serve as useful auxiliary instruments under controlled conditions, but they cannot yet replace validated psychometric tools or clinical judgment in high-stakes assessment. The principal unresolved issues include prompt sensitivity, limited causal reasoning, and insufficient cross-cultural validation. Sections~\ref{ch:03}--\ref{ch:06} examine how current methods attempt to address these limitations in practice.

\section{Psychometric Paradigms and Methods}
\label{ch:03}

Having established that LLMs exhibit both mental-state understanding (Theory of Mind) and role-simulation capabilities (as virtual subjects), a natural question arises: How can these capabilities be systematically applied to measure human psychological constructs? This section examines the methodological frameworks, measurement paradigms, and datasets that operationalize LLM-based psychological assessment.


LLMs introduce new possibilities for psychological measurement. By leveraging their demonstrated capacities for language understanding and role simulation, LLMs can serve as both assessment instruments and analytical engines. Unlike traditional methods constrained by fixed item sets and manual administration, LLM-based approaches enable dynamic interaction, large-scale automated analysis, and integration of multimodal data sources~\cite{ke2025exploring}. However, these advantages come with new methodological considerations: How do we design prompts that reliably elicit psychological information? How do we validate measurements obtained from model outputs? What datasets are appropriate for training and evaluation?

To address these questions, current research has converged on three major measurement paradigms, each operationalizing the LLM's capabilities in distinct ways:

\begin{itemize}
    \item \textbf{Active and Conversational Assessment:} Engaging participants in interactive dialogues, simulating traditional interviews and question–answer assessments.
    
    \item \textbf{Passive Assessment from Natural Language:} Automatically inferring psychological states and traits from naturally occurring texts such as diaries or social media posts.

    \item \textbf{Multimodal Fusion Assessment:} Integrating language with other modalities, such as speech, facial expressions, and physiological signals, to enhance validity and accuracy in psychological assessment.
\end{itemize}

In the following sections, we provide a detailed discussion of these three paradigms. Table~\ref{tab:dataset_cmp_updated} summarizes representative studies, datasets, data scales, and tasks across the three paradigms.

\begin{table*}
    \centering
    \caption{Summary of studies and their datasets for measurement across different psychometric paradigms.}
    \renewcommand{\arraystretch}{1.2}
    \begin{tabular}{p{0.5cm}p{2.6cm}p{2.9cm}p{0.7cm}p{2.8cm}p{5.8cm}}
        \toprule
         Year & \textbf{Method}& \textbf{Datasets}&\textbf{Scale} &\textbf{Task} & \textbf{Performance Highlight}\\
        \midrule
        \multicolumn{6}{c}{\textit{Active and Conversational Assessment}}\\
        \midrule
        2023 & PsyCoT~\cite{yang-etal-2023-psycot} & Essays~\cite{pennebaker1999linguistic} & 2468 & Personality Assessment & Structures admin via CoT reasoning \\
        2023 & Chain of Empathy~\cite{lee2023chain} & EPITOME~\cite{sharma2020empathy} &  10143  & Emotional Reasoning & Incorporates emotional/therapeutic steps \\
        2023 & Thinking Assistants~\cite{park2023thinking} & Thinking Assistants & 179* & Ambiguity Handling & Identifies and clarifies unclear responses \\
        2024 & WundtGPT~\cite{ren2024wundtgpt} & PsyQA~\cite{sun-etal-2021-psyqa} & 22000 & Response Quality & Fine-tuned on counseling dialogues \\
        2024 & Peters et al.~\cite{peters2024large} & crowdsourced annotated & 600* & Personality Assessment & Correlation up to 0.443 with std. measures \\
        2024 & Affective-NLI~\cite{wen2024affective} & CPED~\cite{chen2022cped} & 11835 & Personality Assessment & Reformulates task as Natural Language Inference \\
        2024 & PsyChat~\cite{qiu2024psychat} & SmileChat & 18824 & Behavior Prediction & Five-module agent architecture for control \\
        2024 & CaiTI~\cite{nie2024llm} & CaiTI & 7000 & Mental Health Evaluation & 37-dimensional screening \& intervention system \\
        2025 & OpenR1-Psy~\cite{hu2025beyond} & OpenR1-Psy & 19302 & Response Quality & Tuned on dialogues with reasoning traces \\
        2025 & GATE~\cite{ICLR2025_c9867d5a} & crowdsourced annotated & 388* & Content Recommendation & Agent-based info-gathering dialogue mgmt. \\
        2025 & Zhang et al.~\cite{zhang2025modeling} & AmbigQA~\cite{min2020ambigqa} & 1960 & Opinion Recognition & Optimizes clarification question quality \\
        \midrule
        \multicolumn{6}{c}{\textit{Passive Assessment from Natural Language}}\\
        \midrule
        2023 & Rao et al.~\cite{rao2023can} & (MBTI data) & (N/A) & Personality Assessment & Reformulates items to mitigate response bias \\
        2024 & Rathje et al.~\cite{rathje2024gpt} & Psychological Construct & 47912 & Sentiment Recognition & Comparable accuracy to fine-tuned models \\
        2024 & Niu et al.~\cite{niu2024text} & Emobank~\cite{buechel-hahn-2017-emobank} & 10000 & Sentiment Recognition & Uses examples to guide classification \\
        2024 & Mental-Alpaca~\cite{xu2024mental} & Dreaddit~\cite{turcan2019dreaddit} & 3553 & Stress Prediction & Instruction-tuned on social media posts \\
        2024 & Hu et al.~\cite{hu2024llm} & (Social media posts) & (N/A) & Post analysis & Distills LLM insights into small BERT model \\
        2025 & PostToPersonality~\cite{ma2025post} & MBTI Type Dataset~\cite{kaggle:mbti-type} & 8675 & Personality Assessment & RAG reduces hallucination for rare types \\
        2025 & Teng et al.~\cite{teng2025enhancing} & E-DAIC~\cite{ringeval2019avec} & 275* & Depression Detection & 4-stage reasoning (sentiment, cause, etc.) \\
        2025 & Shah et al.~\cite{shah2025advancing} & Online Depression~\cite{ijcai2017p536} & 40000 & Depression Detection & Incorporates emoji and behavioral signals \\
        \midrule
        \multicolumn{6}{c}{\textit{Multimodal Fusion Assessment}}\\
        \midrule
        2023 & WorkNet~\cite{amadori2023user} & (VR Driving Sim) & (N/A) & Cognitive workload & End-to-end model from physiological signals \\
        2024 & A2II~\cite{feng2024autonomous} & Twitter-2015 \& 2017~\cite{zhang2018adaptive,lu2018visual} & 11310 & Sentiment Recognition & Q-Former fusion; auto-selects modality \\
        2024 & Yang et al.~\cite{yang2024empirical} & Twitter-2015 \& 2017~\cite{zhang2018adaptive,lu2018visual} & 11310 & Sentiment Recognition & Converts images to text for few-shot learning \\
        2024 & Englhardt et al.~\cite{englhardt2024classification} & Globem~\cite{xu2022globem_neurips} & 497* & Mental Health Evaluation & CoT reasoning on wearable data (61.1\% acc) \\
        2025 & ProMind-LLM~\cite{zheng2025promind} & PMData~\cite{thambawita2020pmdata} & 16* & Mental Health Evaluation & Integrates subjective text and objective behavior \\
        \bottomrule
    \end{tabular}
    \label{tab:dataset_cmp_updated}
    \begin{tablenotes}
\small
\item \textit{Note:} ``*'' indicates participant counts; all other numeric entries indicate the total number of data instances.
\end{tablenotes}
\end{table*}

\subsection{Active and Conversational Assessment}
Active and conversational assessment refers to LLMs engaging in real-time dialogue with participants through structured or semi-structured questioning. This paradigm resembles traditional clinical interviews, where the instrument actively elicits responses rather than analyzing existing data. Below we examine the core methods that enable such interactive assessment.

\textbf{Prompting-Based Methods.} The most direct approach involves prompting LLMs to conduct psychological interviews or administer questionnaires without modifying model parameters. Zero-shot prompting provides the model with task descriptions and assessment objectives, relying on pretrained knowledge to generate appropriate questions and interpret responses. Peters et al.~\cite{peters2024large} demonstrated this by instructing GPT-4 to conduct open-ended personality interviews, achieving correlations up to 0.443 with standardized measures. In contrast, few-shot prompting enhances performance by including demonstration examples that guide questioning styles and response interpretations~\cite{zhu2025investigating}. To improve assessment depth, chain-of-thought (CoT) prompting decomposes the process into explicit reasoning stages. PsyCoT~\cite{yang-etal-2023-psycot} structures questionnaire administration as iterative reasoning chains: the model presents an item, interprets the response in relation to the psychological construct, updates its hypothesis, and determines the next question. The Chain of Empathy framework~\cite{lee2023chain} extends this by incorporating emotional reasoning steps, analyzing affective content and considering therapeutic principles before formulating questions. These methods are particularly relevant when assessment objectives include both measurement and participant engagement.

\textbf{Model Adaptation Through Training.} While prompting leverages pretrained capabilities, training-based approaches adapt models more fundamentally to psychological assessment tasks. Instruction tuning trains models on datasets of instruction-response pairs demonstrating desired assessment behaviors. Ren et al.~\cite{ren2024wundtgpt} developed WundtGPT through fine-tuning on the PsyQA dataset~\cite{sun-etal-2021-psyqa} containing 22,000 counseling dialogues, enabling the model to generate questions following clinical interview protocols. Similarly, Hu et al.~\cite{hu2025beyond} created OpenR1-Psy with 19,302 multi-turn dialogues enriched with simulated clinician reasoning processes, allowing models to internalize professional questioning strategies and maintain coherent trajectories across extended interactions. The Affective-NLI framework~\cite{wen2024affective} reformulates personality assessment as natural language inference and fine-tunes models on the CPED dataset~\cite{chen2022cped} containing 11,835 dialogue-personality pairs, improving both accuracy and interpretability. Unlike prompting-based methods, these training methods enable models to handle complex interactive dynamics and recognize subtle psychological cues that may not be detected through prompting alone.

\textbf{Agent-Based Methods and Interactive Strategies.} More sophisticated assessment systems employ agentic architectures where specialized modules coordinate to conduct comprehensive evaluations. PsyChat~\cite{qiu2024psychat} implements a five-module architecture encompassing client behavior recognition, counselor strategy selection, input packing, response generation, and response selection, enabling explicit control and domain knowledge integration. CaiTI~\cite{nie2024llm} extends this into a complete screening and intervention system with modules for 37-dimensional mental health assessment. The GATE framework~\cite{ICLR2025_c9867d5a} actively manages information-gathering dialogues through separate modules for question generation, response interpretation, and preference learning. These systems also incorporate interactive clarification mechanisms to handle ambiguous participant responses. Park et al.~\cite{park2023thinking} introduced reflective questioning strategies where the LLM identifies unclear aspects and formulates targeted follow-up questions. Zhang et al.~\cite{zhang2025modeling} proposed double-turn training that optimizes clarification quality, training models to generate clarifying questions when facing ambiguity before providing refined assessments. Such approaches address limitations of single-turn systems that may misinterpret vague responses.

\textbf{Representative Datasets.} The development of active assessment methods relies on several key datasets. The PsyQA dataset~\cite{sun-etal-2021-psyqa} contains 22,000 multi-turn psychological counseling dialogues between users and counselors, providing rich examples of professional questioning strategies. The Essays dataset~\cite{pennebaker1999linguistic} includes 2,468 text samples annotated with personality traits, commonly used for evaluating personality assessment methods. The CPED dataset~\cite{chen2022cped} provides 11,835 conversational exchanges paired with personality labels, designed specifically for dialogue-based personality recognition. OpenR1-Psy~\cite{hu2025beyond} provides 19,302 counseling dialogues enriched with explicit reasoning traces, enabling models to learn both questioning strategies and the underlying clinical reasoning. These datasets differ in scale, annotation granularity, and target constructs, collectively supporting the development of diverse active assessment approaches.

\subsection{Passive Assessment from Natural Language}
Passive assessment infers psychological states from naturally occurring text, such as social media posts, diary entries, or clinical notes, without active interaction. Unlike active assessment, passive methods must work with whatever textual evidence is available. This section examines methods that adapt LLMs for psychological inference from natural language.

\textbf{Prompting and Reasoning Methods.} The most straightforward approach involves formulating psychological measurement as a text classification task and prompting LLMs to categorize or score input texts. Zero-shot approaches provide construct definitions and classification instructions without training examples. Rathje et al.~\cite{rathje2024gpt} evaluated this across 15 datasets covering 12 languages and four psychological constructs, finding that GPT-4 achieved accuracy comparable to fine-tuned models without task-specific training. Few-shot methods enhance performance by including annotated examples in the prompt context~\cite{niu2024text}. To improve inference accuracy and interpretability, chain-of-thought prompting guides LLMs through explicit reasoning steps that mirror clinical assessment procedures. Teng et al.~\cite{teng2025enhancing} proposed a CoT framework for depression detection that decomposes the task into four sequential stages: sentiment analysis, binary classification, cause identification, and severity assessment. Experiments on the E-DAIC dataset~\cite{ringeval2019avec} showed this structured approach substantially improved detection accuracy and interpretability by revealing which textual features the model considers indicative of psychological states. Task reformulation represents another direction, where assessment problems are recast into formats more amenable to LLM capabilities. Rao et al.~\cite{rao2023can} addressed systematic biases in MBTI assessment by reformulating questionnaire items as neutral statements rather than direct personality queries, mitigating response biases and improving validity.

\textbf{Knowledge Enhancement and Model Adaptation.} To address the hallucination problem and improve reliability, retrieval-augmented generation (RAG) grounds model outputs in verifiable external knowledge. Ma et al.~\cite{ma2025post} developed the PostToPersonality framework that retrieves relevant psychological literature and annotated examples similar to the input text, conditioning predictions on both the input and retrieved context. Experiments on the MBTI Type Dataset~\cite{kaggle:mbti-type} containing 8,675 profiles demonstrated that RAG substantially reduced hallucination rates and improved accuracy for underrepresented personality types. Instruction fine-tuning adapts models specifically for psychological inference tasks. Xu et al.~\cite{xu2024mental} developed Mental-Alpaca by instruction-tuning LLaMA on the Dreaddit dataset~\cite{turcan2019dreaddit} containing 3,553 social media posts, achieving substantial improvements over zero-shot and few-shot configurations and approaching specialized supervised models. Shah et al.~\cite{shah2025advancing} extended this by fine-tuning models on depression detection data incorporating emoji-based emotional features and behavioral signals from 40,000 posts. Fusion-distillation methods address computational constraints by using LLMs to generate rich intermediate representations distilled into smaller models. Hu et al.~\cite{hu2024llm} proposed a framework where GPT-4 analyzes posts from semantic, emotional, and linguistic perspectives, incorporating these analyses into a contrastive learning framework that trains a compact BERT-based model.

\textbf{Representative Datasets.} Passive assessment methods rely on diverse datasets reflecting natural language contexts. The Emobank dataset~\cite{buechel-hahn-2017-emobank} contains 10,000 English sentences annotated with dimensional emotion scores, widely used for emotion recognition research. The Dreaddit dataset~\cite{turcan2019dreaddit} provides 3,553 Reddit posts labeled for stress, capturing authentic expressions of psychological distress in online communities. The E-DAIC dataset~\cite{ringeval2019avec} includes clinical interview transcripts from 275 participants, offering high-quality clinical material for depression assessment. The MBTI Type Dataset~\cite{kaggle:mbti-type} aggregates 8,675 social media profiles with self-reported personality types, enabling large-scale personality inference studies. The Pandora dataset~\cite{gjurkovic-etal-2021-pandora} contains 9,067 texts annotated with multiple psychological dimensions, supporting multi-construct assessment research. These datasets vary in source domain, annotation quality, and psychological constructs, collectively enabling the development and evaluation of passive assessment methods.

\subsection{Multimodal Fusion Assessment}
Multimodal fusion assessment integrates text with other modalities, including acoustic signals, facial expressions, and behavioral data, to provide more comprehensive psychological evaluations. The core principle is to approximate clinicians’ multi-channel perception, which considers both verbal and non-verbal cues.

Much of the current research focuses on combining textual and visual information. For example, the Multimodal Aspect-Oriented Sentiment Classification (MABSC) task requires models to determine people’s sentiment toward specific entities (e.g., a celebrity or a product) by jointly analyzing text and images. However, existing approaches face two major challenges: (1) key image information is often lost during fusion, and models trained on small datasets have weak fusion capabilities; (2) irrelevant images can interfere with model predictions. To address these issues, Feng et al.~\cite{feng2024autonomous} proposed the multimodal model A\textsuperscript{2}II, which leverages a Q-Former module to efficiently link large vision-language models, improving fusion performance without significantly increasing parameter size. It also incorporates an automatic selector that adaptively switches between “text-only” and “multimodal” decision-making based on image relevance, thereby mitigating interference from irrelevant visual data. Yang et al.~\cite{yang2024empirical} further explored leveraging ChatGPT’s in-context learning capabilities to tackle the Multimodal Entity-Based Sentiment Analysis (MEBSA) task, aiming to reduce dependence on large-scale labeled data. By converting images into descriptive text and combining them with original text inputs, Li designed zero-shot and few-shot instruction-learning strategies. Additionally, an Entity-Aware Contrastive Learning model was developed to retrieve semantically similar samples, thereby enhancing few-shot performance. Building on these pipeline-style strategies, native multimodal LLMs such as Gemini, GPT-4V/GPT-4o, and Qwen2-VL further shift ``fusion'' from explicit feature aggregation to end-to-end cross-modal interpretation~\cite{reid2024gemini,google2025gemini2,wang2024qwen2vl}. In particular, Lian et al.~\cite{lian2023gpt4vemotion} used GPT-4V as a zero-shot benchmark for multimodal emotion understanding, showing that audio-visual emotional cues can be interpreted more directly without relying on separate facial expression recognition modules. For psychological assessment, this better matches clinical observation by jointly considering what a participant says, how it is said, and the accompanying nonverbal context, while also introducing new evaluation requirements such as temporal grounding and modality consistency.

Beyond text-image fusion, researchers have introduced objective behavioral data to compensate for the subjective bias inherent in self-reported textual records. Englhardt et al.~\cite{englhardt2024classification} investigated the use of LLMs in analyzing behavioral health data collected from smartphones and wearable devices, such as activity levels, sleep patterns, and social interactions. They developed a CoT prompting method that enables LLMs to assess the relationship between these behavioral indicators and mental health conditions like depression and anxiety, and generate specific reasoning processes. Subsequently, LLM was prompted to perform a binary classification task, achieving an accuracy rate as high as 61.1\%. The generated textual summaries during the reasoning processes provided valuable support for clinical decision-making. Zheng et al.~\cite{zheng2025promind} proposed ProMind-LLM, a psychological risk assessment model that integrates subjective records with objective behavioral data such as heart rate and sleep quality. ProMind-LLM was first trained on 100,000 domain-specific mental health articles, enhanced with counterfactual data to improve domain robustness, and optimized through behavioral data structuring for better LLM comprehension. By implementing causal reasoning into CoT framework, the model further improved the accuracy and interpretability of mental health risk predictions. Beyond these text- and behavior-oriented methods, Amadori and Demiris~\cite{amadori2023user} introduced WorkNet, an end-to-end sequential deep learning model for cognitive workload estimation from multimodal physiological signals in a virtual reality driving simulator.

Overall, multimodal fusion enables LLMs to extract psychological cues from diverse channels, improving the accuracy and reliability of assessments beyond what single-modality analysis can achieve.

\subsection{Comparison with Traditional Approaches}
While these three LLM-based paradigms offer novel capabilities, they also have limitations compared to traditional approaches. Below we compare them in terms of reliability and interpretability.

\textbf{Reliability.} Traditional psychological assessment tools, such as standardized questionnaires and structured interviews, have undergone extensive scientific validation over decades, providing well-established statistical metrics and theoretical grounding. These methods are widely regarded as convincing tools in psychological assessment due to their high reliability~\cite{miller2001inpatient}. In contrast, LLM-based methods remain in an early stage of development. Although numerous studies have demonstrated promising performance~\cite{yang-etal-2023-psycot, shah2025advancing, zheng2025promind}, their accuracy heavily depends on both model capability and prompt design. LLMs are also prone to hallucinations, producing responses that appear plausible but lack factual grounding~\cite{huang2025survey}. When hallucinations occur in clinical decision-making, they may lead to misdiagnosis and potentially severe consequences. Future research should focus on mitigating hallucinations to further enhance the reliability of LLM-based assessments and facilitate their safe application in real-world settings.

\textbf{Interpretability.} Traditional methods also maintain a distinct advantage in this dimension. Questionnaire items typically correspond directly to specific symptoms or theoretical constructs, making them easy for both clinicians and patients to understand and track~\cite{gilbert2015use}. Similarly, structured interviews allow direct mapping between dialogue content and symptom presentation~\cite{brinkmann201414}. In contrast, while some studies have attempted to improve LLM interpretability by revealing step-by-step reasoning~\cite{qiu2024psychat, teng2025enhancing}, such outputs remain inherently opaque and difficult to audit. This black-box nature poses a major barrier to clinical adoption. Techniques such as retrieval augmented generation (RAG) could improve interpretability by grounding outputs in verifiable evidence~\cite{kermani2025systematic}. Beyond technical solutions, evaluations of LLM-based approaches should not focus solely on accuracy metrics; they should also assess text generation quality, including fidelity to the source material, coherence, and logical consistency.


In summary, traditional methods retain clear advantages in reliability and interpretability, while LLM-based methods excel in automation and cost-efficiency. Future research should focus on developing hybrid paradigms that integrate LLM-driven automation with the robust validation and transparency of traditional approaches. Such integration could improve overall efficiency while maintaining the scientific rigor and clinical trustworthiness of assessment results.

\section{Personality Trait Assessment}
\label{ch:04}
Building on the methodological paradigms discussed above, this section focuses on personality trait assessment. Section~\ref{sec:standardized_scales} reviews assessments with standardized scales and interactive paradigms. Section~\ref{sec:social_media} examines personality profiling from social media data, and Section~\ref{sec:complex_personality} explores the assessment of complex personality traits.

\begin{table*}[!t]
\centering
\begin{threeparttable}
\caption{Comparison of Classic Personality Assessment Frameworks. Note that frameworks such as Eysenck's PEN model, Cattell's 16PF, MMPI-2, CPI, and HPI remain underrepresented or entirely absent from LLM personality assessment studies.}
\label{tab:personality_frameworks}
\small
\setlength{\tabcolsep}{3pt}
\renewcommand{\arraystretch}{1.15}
\begin{tabular}{@{}>{\raggedright\arraybackslash}p{2.05cm} >{\raggedright\arraybackslash}p{4.15cm} >{\raggedright\arraybackslash}p{3.45cm} >{\raggedright\arraybackslash}p{2.05cm} >{\raggedright\arraybackslash}p{3.95cm}@{}}
\toprule
\textbf{Framework} & \textbf{Dimension Characteristics} & \textbf{Assessment Focus}  & \textbf{Validation Status} & \textbf{LLM Applications} \\
\midrule
\textbf{Big Five} \cite{goldberg1993structure} & Openness, Conscientiousness, Extraversion, Agreeableness, Neuroticism & Broad personality variation across cultures  & High  & \cite{serapio2023personality, yan2024predicting, salecha2024large, shum2025big, cohen2025exploring, hartley2025personality} \\
\midrule
\textbf{HEXACO} \cite{ashton2007empirical} & Honesty-Humility, Emotionality, eXtraversion, Agreeableness, Conscientiousness, Openness & Moral character and interpersonal behavior & Moderate-High & \cite{bodrovza2024personality, barua2024psychology, ren2024valuebench, wang2025exploring, zheng2025lmlpa} \\
\midrule
\textbf{MBTI} \cite{myers2003mbti} & E/I, S/N, T/F, J/P dichotomies & Cognitive preferences and career counseling & Low  & \cite{lu2023illuminating,zhou2024chinese,ma2025post, wang2025mbti, li2025can} \\
\midrule
\textbf{Dark Triad} \cite{paulhus2002dark} & Narcissism, Machiavellianism, Psychopathy & Maladaptive interpersonal traits & Moderate & \cite{li-etal-2024-evaluating-psychological, lu2023illuminating, tu4965442using, lee-etal-2025-llms} \\
\midrule
\textbf{Eysenck's PEN} \cite{eysenck1991dimensions} & Psychoticism, Extraversion, Neuroticism & Biological temperament and personality structure & Moderate & N/A \\
\midrule
\textbf{Cattell's 16PF} \cite{cattell2008sixteen} & 16 primary factors (e.g., Warmth, Reasoning, Emotional Stability) & Comprehensive personality profiling & Moderate & \cite{chittem2025sac} \\
\midrule
\textbf{MMPI-2} \cite{butcher2010minnesota} & 10 clinical scales + validity scales & Psychopathology and clinical diagnosis & High & N/A \\
\midrule
\textbf{CPI} \cite{gough1956california} & 20 scales measuring social behavior & Normal personality and interpersonal effectiveness & Moderate-High & N/A \\
\midrule
\textbf{HPI} \cite{hogan1992hogan} & Adjustment, Ambition, Sociability, Interpersonal Sensitivity, Prudence, Inquisitive, Learning Approach & "Bright-side" personality for leadership & High & N/A \\

\bottomrule
\end{tabular}
\begin{tablenotes}
\small
\item \textit{Note:} 16PF = 16 Personality Factors; MMPI-2 = Minnesota Multiphasic Personality Inventory-2; CPI = California Psychological Inventory; HPI = Hogan Personality Inventory. Validation status reflects psychometric rigor and empirical support in personality assessment literature. The absence of MMPI-2, CPI, and HPI from LLM studies is largely attributable to copyright and licensing restrictions that prohibit the reproduction of test items in training data or prompts.
\end{tablenotes}
\end{threeparttable}
\end{table*}

\subsection{Assessing Personality through Interactive Paradigms}
\label{sec:standardized_scales}
Personality traits are enduring patterns of thought, emotion, and behavior. Several frameworks describe and quantify personality, including the Big Five~\cite{goldberg2013alternative}, MBTI~\cite{myers1962myers, myers2003mbti}, and HEXACO~\cite{ashton2007empirical}. The Big Five model (Openness, Conscientiousness, Extraversion, Agreeableness, and Neuroticism) is widely used in contemporary psychology. HEXACO adds Honesty-Humility. These traits have traditionally been measured through self-report questionnaires such as BFI~\cite{john1991big} and HEXACO-60~\cite{ashton2009hexaco}. Researchers have increasingly explored whether LLMs can infer personality traits from text~\cite{peters2024large}.

Early studies relied on manually constructed linguistic feature libraries, such as the Linguistic Inquiry and Word Count (LIWC) dictionary, to extract lexical and syntactic features from text, which were then fed into machine learning models for personality prediction. Park et al.~\cite{park2015automatic} combined LIWC and NRC-based features, achieving 61\% accuracy in predicting openness\cite{maharjan2025psychometric}. Despite these promising results, such approaches demand extensive feature engineering, struggle to capture deeper semantic patterns, and often exhibit limited generalizability across contexts.
The advent of LLMs has shifted this paradigm. LLMs can directly infer behavioral tendencies and personality traits from raw text without handcrafted features. Wang et al.\cite{wang2025evaluating} applied GPT-4 to predict human responses on BFI items using natural language inputs, demonstrating strong convergent and discriminant validity relative to traditional self-report scores. 

Beyond extending traditional survey-based approaches, novel paradigms have emerged to overcome the static limitations of questionnaire methods. The PsychoGAT framework~\cite{yang2024psychogat}  converts survey items into interactive narrative scenarios, where participants’ behavioral choices are used to infer personality traits, enhancing both reliability and participant engagement. Li et al.~\cite{li2025traits} employ psychology-informed prompts to extract personality-relevant semantic features from text and fuse these with multimodal signals (e.g., audio and video) to achieve a more robust and holistic assessment. 

These methods offer automated and scalable approaches to personality assessment that complement traditional inventories. By capturing subtle textual and latent cues, LLM-based methods extend personality research into contexts where conventional questionnaires are impractical or infeasible.

\subsection{Personality Profiling from Social Media}
\label{sec:social_media}
Social media generates large volumes of unstructured text through user interactions. These naturally occurring texts provide a basis for inferring personality traits without self-report surveys~\cite{pramod2018identifying}. Compared with traditional NLP approaches based on word frequency or shallow features, LLMs leverage deeper semantic understanding to uncover psychological and behavioral patterns~\cite{tong2024advirds}.

Several large-scale social media datasets have been developed to support personality research. The CMACD dataset~\cite{zhou2024chinese} integrates text, MBTI types, and emotional intensity scores from more than 11,000 Weibo users, offering resources for personality and affective studies in Chinese contexts. The PANDORA dataset~\cite{gjurkovic-etal-2021-pandora} contains approximately 17 million Reddit comments annotated with Big Five, MBTI, and Enneagram labels, serving as a benchmark for personality prediction in English social media. The MuMiN dataset~\cite{nielsen2022mumin}, originally designed for misinformation detection, contains tweets, user metadata, and heterogeneous graph relations that can support multimodal personality analysis. These cross-lingual and multimodal resources enable evaluation of model robustness across different cultural and contextual settings.

Methodological developments in LLM-based personality analysis can generally be grouped into three categories.

\textbf{Prompt-based inference.} This approach relies on general knowledge in LLMs to predict traits directly from text using carefully designed prompts. Peters et al.~\cite{peters2024large} employed GPT-3.5 and GPT-4 to perform zero-shot predictions of Big Five traits from Facebook status updates. The results indicated that the models achieved correlations comparable to traditional methods, although predictive accuracy varied between demographic groups. 
Yang et al.~\cite{yang-etal-2023-psycot} proposed PsyCoT, which reformulates questionnaire items into step-by-step chains of thought, guiding GPT-3.5 to reason in a structured manner. This approach produces more stable and consistent trait scores relative to direct prompting. Ji et al.~\cite{ji2023chatgpt} compared different prompting strategies and found that zero-shot chain-of-thought improved ChatGPT's accuracy while generating natural-language explanations that enhance transparency.

\textbf{Feature-based modeling with LLM embeddings.} In this approach, LLMs are used as feature extractors to transform text into high-dimensional semantic embeddings, which are then processed by lightweight supervised models for personality prediction. Maharjan et al.~\cite{maharjan2025psychometric} tested various embeddings on the PANDORA dataset and reported consistent gains over prompt-based methods in both accuracy and psychometric validity, with further improvements observed when combined with traditional linguistic features.

\textbf{Domain-specific fine-tuning.} This approach involves adjusting model parameters on specialized datasets to improve sensitivity to personality-related signals. Wang et al.~\cite{wang2024continuous} fine-tuned RoBERTa-base on the PANDORA dataset to generate continuous Big Five scores, significantly outperforming binary classification models. Similarly, Li et al.~\cite{li-etal-2025-big5} introduced BIG5-CHAT, training LLMs on conversational datasets to enhance personality expression in dialogue. Their results showed stronger correlations with psychometric dimensions and cognitive task performance, underscoring the effectiveness of fine-tuning.

Social media personality profiling extends the applicability of LLMs beyond questionnaire-based simulations by incorporating real-world behavioral data, which improves ecological validity. Table~\ref{tab:performance_trends} summarizes the evolution of key methods, metrics, and performance benchmarks in LLM-based personality assessment. The table shows a methodological shift from feature engineering to end-to-end deep learning~\cite{wang2024continuous}, refinement of evaluation metrics toward psychometrically grounded measures such as reliability and correlation~\cite{peters2024large, maharjan2025psychometric}, and steady performance improvements across diverse tasks. These reflect growing sophistication in capturing human personality from text with improved psychological validity.


\begin{table*}[ht]
\centering
\caption{Performance trends of LLM-based personality profiling methods across years}
\label{tab:performance_trends}
\renewcommand{\arraystretch}{1.2}
\begin{tabular}{p{0.7cm} >{\raggedright\arraybackslash}p{3.5cm} >{\raggedright\arraybackslash}p{3.0cm} >{\raggedright\arraybackslash}p{2.5cm} >{\raggedright\arraybackslash}p{5.5cm}}
\toprule
\textbf{Year} & \textbf{Method} & \textbf{Dataset} & \textbf{Metric} & \textbf{Performance} \\
\midrule
2015 & LIWC + NRC \cite{park2015automatic} & Big-5 & Accuracy & 61\% (Openness) \\[2pt]
2021 & BERT + Softmax \cite{personality2021bert} & Big-5 & Accuracy & 75.8\% \\[2pt]
2022 & AWS-EP (Multi-task) \cite{awsep2022} & Big-5 & MSE & 564.12 \\[2pt]
\midrule
2023 & PsyCoT \cite{yang-etal-2023-psycot} & Essays & F1-score & 0.5843 \\[2pt]
2023 & ChatGPT Zero-Shot CoT \cite{ji2023chatgpt} & PAN & AIP & +2.9\% accuracy improvement \\[2pt]
2023 & BPD Simulation \cite{hadar2023plasticity} & EAS & Affective Richness & Stronger and complex affective responses \\[2pt]
2023 & Combined Assessment \cite{lu2023illuminating} & MBTI, Big Five, SD-3 & Trait Scores & Consistent scores across normative and dark dimensions \\
2023 & Narcissism Scoring \cite{tu4965442using} & 1,669 open-ended responses & Agreement & Strong agreement with human experts \\
\midrule
2024 & RoBERTa fine-tuned \cite{wang2024continuous} & PANDORA & $R^2$ & 0.59 (Highest) \\[2pt]
2024 & GPT-4 Zero-shot \cite{peters2024large} & 566 US participants & Correlation (r) & Mean r=0.443 \\[2pt]
2024 & PsychoGAT (GPT-4) \cite{yang2024psychogat} & GPT-4 Generated & Human & Cronbach's $\alpha$ \& 0.97 (Extroversion) \\[2pt]
2024 & SVD-based latent traits \cite{suh2024rediscovering} & PersonaLLM & Variance Explained & 74.3\% (Top 5 factors) \\[2pt]
2024 & Dark Triad (GPT-4) \cite{li-etal-2024-evaluating-psychological} & SD-3 questionnaire & SD-3 Scores & Higher on Machiavellianism \& Narcissism \\
2024 & Dynamic Shifts \cite{song2024identifying} & MBTI & Accuracy & 93.5\%\\
\midrule
2025 & Psychometric Embeddings \cite{maharjan2025psychometric} & PANDORA & Cronbach's $\alpha$; MSE & 0.63; 526.9 \\[2pt]
2025 & BIG5-CHAT \cite{li-etal-2025-big5} & Human-annotated Big-5 & Trait Correlation & Closer to human than prompts \\[2pt]
2025 & LLM Consistency Framework \cite{lee-etal-2025-llms} & TRAIT & Trait Correlation & Unconventional trait patterns \\[2pt]
2025 & Generative Implicit Bias \cite{ye2025generative} & Value-Laden Perception & Cohen's d & 0.42 (Gender-Career) \\[2pt]
2025 & GPT-4 BFI Prediction \cite{wang2025evaluating} & Essays & SR/ROR Correlation & 0.64 \\
2025 & Multimodal Fusion \cite{li2025traits} & AVI 2025 & MSE & 0.1095 \\
\bottomrule
\end{tabular}
\end{table*}

\subsection{Assessing Complex Personality Disorders}
\label{sec:complex_personality}
In addition to normative traits such as the Big Five, complex personality disorders, including the Dark Triad (narcissism, Machiavellianism, and psychopathy)~\cite{paulhus2002dark} and borderline personality disorder (BPD), have substantial clinical relevance and broader social implications.

Before examining how LLMs assess human personality disorders, researchers have first assessed the intrinsic personality tendencies of LLMs using established psychometric instruments. Li et al.~\cite{li-etal-2024-evaluating-psychological} applied the Short Dark Triad (SD-3) scale to GPT models and found that, despite safety alignment measures, the models exhibited higher scores than human averages in Machiavellianism and narcissism, suggesting latent dark traits. Similarly, Lu et al.~\cite{lu2023illuminating} combined MBTI, Big Five, and SD-3 assessments and demonstrated that LLMs can be consistently evaluated across both normative and dark dimensions. Tu et al.~\cite{tu4965442using} further evaluated ChatGPT’s scoring of open-ended responses to narcissism-related items from 1,669 participants and found strong agreement with human experts. This suggests that LLMs may reliably detect and quantify narcissistic tendencies, providing a basis for analyzing other dark personality traits. Extending this multidimensional approach, Lee et al.~\cite{lee-etal-2025-llms} developed a multidimensional personality assessment framework covering both Dark Triad and Big Five dimensions, confirming consistent dark trait patterns and underscoring the significant influence of training data on model behavior.


Beyond trait-specific evaluations, researchers have begun to assess complex personality through dynamic, multidimensional, and implicit dimensions~\cite{ye2025large}. First, LLMs exhibit dynamic personality shifts across contexts: conscientiousness often increases in work-related tasks, while agreeableness tends to rise in emotional-support settings~\cite{song2024identifying}. This suggests that personality in LLMs is not a static attribute but a fluid construct that responds to situational demands. Second, analyses of multidimensional trait interactions reveal that the internal personality structures of LLMs can diverge from human norms. Studies have reported unconventional correlations between traits, such as a stronger negative association between openness and conscientiousness than is typically observed in humans~\cite{suhr2023challenging, huang2023revisiting}. Third, research into implicit personality and bias has revealed latent tendencies that are not apparent through direct assessment. Methods adapted from the Implicit Association Test (IAT) have revealed latent biases, such as a male–science association, even when models explicitly deny holding prejudices~\cite{wen2024evaluating, zhao2025explicit}. Social desirability bias also emerges, where undesirable traits are suppressed in direct assessments but surface during indirect or projective tasks, revealing a gap between explicit self-presentation and implicit inclinations~\cite{ye2025generative, biedma2024beyond}. These perspectives broaden the scope of complex personality research from surface-level detection toward a deeper understanding of LLMs' implicit and emergent psychological properties.

\textbf{Psychometric Summary.} Among psychometric properties, convergent validity has received the most empirical support: LLM-derived Big Five scores show moderate-to-strong correlations with self-report measures~\cite{peters2024large,wang2025evaluating}. Internal consistency has been demonstrated for interactive paradigms such as PsychoGAT ($\alpha=0.97$)~\cite{yang2024psychogat}. Serapio-Garc{\'\i}a et al.~\cite{serapio2025psychometric} proposed a comprehensive psychometric framework for evaluating personality in LLMs, showing that large instruction-tuned models yield reliable measurements and that specific personality profiles can be shaped for downstream tasks. However, test-retest reliability, discriminant validity across closely related constructs, and criterion validity (predicting real-world outcomes) remain largely untested. Cross-cultural measurement invariance is also an open question, as most studies rely on English-language data from Western populations.

Overall, these studies highlight the potential of LLMs in identifying complex personality traits, suggesting new directions for computational psychology. Nevertheless, systematic psychometric validation is required before clinical application.
\section{Cognitive State and Mental Health Assessment}
\label{ch:05}
This section focuses on mental health applications. Section~\ref{sec:common_mental_health_issues} examines automated detection of psychological disorders. Section~\ref{sec:emotion_and_affective_state} covers emotion and affective state analysis. Section~\ref{sec:psychological_crisis} discusses psychological crisis detection, and Section~\ref{sec:broader_applications} highlights emerging applications in educational, organizational, and social psychology.

\subsection{Automated Detection of Mental Health Issues}
\label{sec:common_mental_health_issues}
Systematic screening and early intervention are crucial in mitigating the global burden of mental health disorders\cite{colizzi2020prevention}. Early detection methods for psychological conditions initially relied on BERT~\cite{devlin2019bert} or its variants for text-based analysis. Verma et al.~\cite{verma2023ai} employed RoBERTa~\cite{liu2019roberta} to detect potential signs of depression by capturing linguistic cues including sentiment patterns, language usage, emotional expressions, and topics discussed. 
 
The advancement of GPT~\cite{yenduri2023generative} has led researchers to apply GPT models to identify linguistic indicators in text that may reflect underlying psychological disorders. LLMs have shown significant potential in detecting common psychological conditions based on text analysis, such as depression, anxiety, and post-traumatic stress disorder (PTSD). Kuzmin et al.~\cite{kuzmin2024mental} compared the effectiveness of traditional machine learning approaches, encoder-based models, and LLMs in detecting depression and anxiety. The experimental results demonstrate that LLMs consistently surpass traditional approaches, especially in scenarios involving noisy and limited data, where textual inputs vary considerably in length and genre. Xu et al.~\cite{xu2024mental} conducted a comprehensive evaluation of multiple LLMs on a variety of mental health prediction tasks using online text data. The study encompasses a broad experimental design, systematically exploring zero-shot prompting, few-shot prompting, and instruction fine-tuning. Results demonstrate that instruction fine-tuning substantially enhances LLM performance across all tasks simultaneously. Their best-performing fine-tuned models, Mental-Alpaca and Mental-FLAN-T5, achieve performance comparable to state-of-the-art task-specific language models.
 
Recent studies have integrated multimodal data, such as speech, facial expressions, and physiological signals, to achieve more comprehensive mental health assessments. Danner et al.~\cite{danner2023advancing} utilized multimodal data, including text, audio, and video from the DAIC datasets, to assess depression severity and generate transcriptions. 
 
Utilizing LLMs as virtual agents to engage in conversational interactions with users has been demonstrated to be effective in assessing and analyzing mental health states. Tao et al.~\cite{tao2023classifying} proposed a virtual interaction framework powered by LLMs, enabling participants to engage in dialogue with a virtual character. The framework leverages enhanced LLMs to analyze mental health concerns in real time and provides supportive suggestions during the conversation to help alleviate users' current psychological distress. 
 
LLMs can also be integrated as modular components in multi-stage analytical pipelines, enhancing the identification and assessment of psychological conditions. Sadeghi et al.~\cite{sadeghi2023exploring} leveraged Whisper~\cite{radford2023robust} to transcribe speech, GPT-3.5-Turbo to summarize textual content, and a text-encoder to extract depression-related features, forming a multi-stage pipeline to predict the PHQ-8 scores from textual data. Qin et al.~\cite{qin2025explainable} introduced an interpretable and interactive depression detection system that leverages LLMs. This system allows users to engage in natural language conversations, enabling a more personalized and nuanced understanding of their mental state through analysis of their social media content. 

LLMs offer scalable tools for early screening and continuous monitoring, particularly in resource-limited settings. As models improve in contextual understanding and multimodal integration, their role in supporting mental health care may expand further.

\subsection{Emotion and Affective State Analysis}
\label{sec:emotion_and_affective_state}
Prior to LLMs, BERT has been widely employed in emotion analysis applications. Stigall et al.~\cite{stigall2024large} found that EmoBERTTiny surpasses baseline models such as BERT-Base-Cased and BERT-Tiny, showing the advantages of task-specific fine-tuning in model performance. 

Building on these BERT-based foundations, LLMs have been applied to sentiment analysis and emotion classification~\cite{stigall2024large, lossio2024comparison}. In sentiment analysis, LLMs categorize text into broad polarity classes, typically positive, neutral, negative, and occasionally mixed, to capture overall emotional tone. In emotion classification, they assign fine-grained emotional labels such as ``joy'', ``sadness'',  ``anger'', and ``fear''~\cite{stigall2024large}, enabling a more nuanced understanding of affective states in textual data~\cite{nandwani2021review}. By detecting affective states in everyday contexts, LLMs may facilitate timely intervention before disorders develop.

Researchers have also enhanced the capacity of LLMs to recognize and interpret emotions by optimizing performance at multiple stages of the training pipeline. Dutta et al.~\cite{dutta2025llm} proposed a method for pretraining a text-based emotion recognition model using unsupervised speech transcripts guided by LLMs. A text-based LLM generates pseudo-labels for the transcripts, which are then used to train an utterance-level emotion recognition model. This LLM-guided labeling approach enables supervision in the absence of annotated emotional labels. Liu et al.\cite{liu2024emollms} proposed EmoLLMs, a series of open-source instruction-following LLMs designed for affective analysis. These models were developed by fine-tuning pre-trained LLMs on the Affective Analysis Instruction Dataset (AAID), a dataset constructed to support emotion and sentiment understanding. Li et al.~\cite{li2025emoverse} proposed Emotion Universe (EmoVerse), a multimodal large language model (MLLM) trained via a Multistage Multitask Sentiment and Emotion (M2SE) instruction tuning strategy. This training strategy enables EmoVerse to recognize affective states and perform reasoning about their underlying causes, enhancing its capacity for nuanced affective understanding. 

In the inference stage, the emotion recognition performance of LLMs can be effectively enhanced through prompt engineering. Hong et al.~\cite{hong2025aer} designed zero-shot and few-shot prompting strategies, incorporating prior dialogue context to enhance ambiguous emotion recognition. Their results demonstrate that LLMs achieve high effectiveness in identifying less ambiguous emotions and show promising potential in recognizing more nuanced, context-dependent emotional states. Li et al.~\cite{li2025revise} introduced a Revise-Reason-Recognize prompting framework designed to enhance the robustness of LLM-based emotion recognition from spoken language in the presence of ASR errors. Their experimental results validate the effectiveness of emotion-specific prompting strategies, ASR error correction mechanisms, and tailored LLM training approaches in improving emotion recognition performance.

Overall, LLMs show potential in emotion recognition by capturing nuanced linguistic and contextual cues. However, challenges in interpretability, bias mitigation, and computational efficiency remain, and further work is needed before reliable deployment in clinical and real-world settings.

\subsection{Psychological Crisis Alert Systems}
\label{sec:psychological_crisis}
Diniz et al.~\cite{diniz2022boamente} demonstrated the strong performance of the BERTimbau Large model in detecting suicidal ideation within a Portuguese-language context. Metzler et al.~\cite{lamsal2024crisistransformers} evaluated the BERT model and found that it correctly classified most of the tweets as either suicidal or off-topic, demonstrating performance comparable to that of human analysts and other state-of-the-art models. Wu et al.~\cite{wu2025psychological} introduced a multi-level framework that employs transfer learning on BERT and integrates domain mental health knowledge, sentiment analysis, as well as behavior prediction modeling techniques. The proposed model is superior to the traditional method in crisis detection accuracy and demonstrates a greater sensitivity to underlying differences in context and emotion. 

LLMs have been applied to detect critical psychological crisis signals. Deng et al.~\cite{deng2025evaluating} found that LLMs perform well on suicidal ideation detection, suicide plan identification, and risk assessment, with gains from few-shot prompting and fine-tuning. Ghanadian et al.~\cite{ghanadian2024socially} proposed an approach that uses generative AI models, such as ChatGPT, Flan-T5, and Llama, to generate synthetic data for the detection of suicidal ideation. This data generation strategy is grounded in social factors derived from psychological literature, aiming to ensure coverage of key elements associated with suicidal thoughts. By incorporating empirically supported risk factors, the method enhances the representativeness and relevance of the synthetic data, which may support more robust and generalizable detection models. Xu et al.~\cite{xu2024utilizing} proposed a system using LLMs for the detection of suicidal tendencies based on social media content. The system integrates prompt engineering and Retrieval-Augmented Generation (RAG)~\cite{gao2023retrieval} techniques to improve detection accuracy, comprising a knowledge retrieval-enhanced module and a judgment module, both using a fine-tuned LLM. Wang et al.~\cite{wang2025multi} proposed a multi-stage framework based on a large language model to improve the extraction of suicide-related social determinants of health (SDoH) from unstructured text. The framework comprises two intermediate stages, context retrieval and relevance verification, followed by a final decision-making stage dedicated to the identification and extraction of SDoH factors. Gao et al.~\cite{gao2025leveraging} employed LLMs for suicide risk detection based on spontaneous speech. They used LLMs as tools for feature extraction in conjunction with traditional acoustic and semantic features. Their findings suggest potential for LLM-based approaches in analyzing speech for suicide risk assessment. 

LLMs have shown performance comparable to traditional screening methods and can process unstructured text at scale for real-time monitoring. However, challenges in interpretability, cultural bias, and data privacy remain. Responsible deployment requires robust frameworks for clinical and community settings.

\textbf{Psychometric Summary.} For mental health detection, concurrent validity has been partially supported: LLM-based methods achieve accuracy comparable to supervised baselines on depression and anxiety screening tasks~\cite{xu2024mental,kuzmin2024mental}. Sensitivity to linguistic markers of distress has been demonstrated across multiple datasets. Badawi et al.~\cite{badawi2025trust_mental} introduced large-scale benchmarks (MentalBench-100k and MentalAlign-70k) that compared LLM judges with human experts across 70,000 ratings, revealing that cognitive attributes such as guidance achieve reliable inter-rater agreement, while affective dimensions like empathy show high point estimates masking large uncertainty, and safety-critical assessments remain unreliable without human oversight. These findings highlight that predictive validity, diagnostic specificity, and inter-rater reliability across different LLMs remain underexplored. Most evaluation still relies on binary classification metrics rather than psychometrically grounded measures such as sensitivity and specificity at clinically meaningful thresholds.

\subsection{Broader Emerging Applications}
\label{sec:broader_applications}
Beyond personality and mental health assessment, psychology LLMs are beginning to support a wider range of application domains. In educational psychology, LLMs have been explored for adaptive teaching and learning-style-aware instruction, suggesting potential for assessing how students understand and engage with different forms of explanation~\cite{weijers-etal-2024-quantifying}. In organizational psychology, LLM-based simulations have been used to study how personality and model capability affect negotiation behavior, providing a scalable testbed for team interaction and workplace decision-making research~\cite{cohen2025exploring}. In social psychology, LLMs have been used both to simulate human samples for behavioral research~\cite{argyle2023out} and to analyze how group-level conventions and collective biases emerge in multi-agent populations~\cite{ashery2025emergent}. These studies indicate that psychology LLMs are expanding from individual trait or symptom assessment toward broader investigations of learning, collaboration, and social behavior.

\begin{table*}[ht]
\centering
\caption{Summary of personality assessment datasets for LLM-based research}
\label{tab:personality_datasets}
\renewcommand{\arraystretch}{1.2}
\begin{tabular}{lllll}
\toprule
\textbf{Dataset} & \textbf{Source} & \textbf{Data Size} & \textbf{Language} & \textbf{Annotation} \\
\midrule
\multicolumn{5}{c}{\textit{Real-World Datasets}}\\
\midrule
\textbf{CMACD} \cite{zhou2024chinese} & Social media (Weibo) & 11k users, 566k posts & Chinese & MBTI and affective labels \\
\textbf{PANDORA} \cite{maharjan2025psychometric} & Social media (Reddit) & 10k users, 17M posts & English & Big Five, MBTI, Enneagram \\
\textbf{PDCH} \cite{cao2025multimodal} & Clinical consultations &  2,937 min audio & Chinese & Audio recording and transcribed text \\

\textbf{EATD} \cite{shen2022automatic} & Semi-structured interviews & 142 participants  & Chinese & Emotional Audio-Textual Depression \\

\textbf{myPersonality} \cite{kosinski2013private} & Social media (Facebook) & 250 users, 9913 posts  & English & Big Five \\
\textbf{WorryWords} \cite{mohammad2024worrywords} & English words & 44,450  & Multi & word–anxiety
associations \\

\textbf{MMPsy} \cite{qin2025mental} & Adolescent volunteers & 7,736  & Chinese & Anxiety and depression detection \\

\midrule
\multicolumn{5}{c}{\textit{Synthetic Datasets}}\\
\midrule

\textbf{PhDGPT} \cite{de2024phdgpt} & GPT-3.5 & 7,736  & English & Depression, Anxiety, and Stress Scale  \\

\textbf{PsychoLexEval} \cite{Abbasi2024PsychoLex} & GPT-4 & 10k  & Persian \& English & Multiple-choice questions  \\

\textbf{CounseLLMe} \cite{de2025introducing} & Claude-3’s Haiku & 400  & Persian \& English & Feelings of conflict and pessimism  \\

\textbf{Bhandari et al.} \cite{bhandari2025can} & GPT-4o & 2020 dialogues  & English & Big Five \\
\textbf{SoulChat} \cite{chen2023soulchat} & ChatGPT & 2M  & Chinese & Dialogue \\
\textbf{Cactus} \cite{lee2024cactus} & GPT-4o & 31,577  & English & CBT dialogue  \\
\textbf{Psych8k} \cite{liu2023chatcounselor} & Hunab and GPT-4 & 8,187  & English & QA Pairs \\

\bottomrule
\end{tabular}
\end{table*}
\section{Technical Challenges and Ethical Considerations}
\label{ch:06}
Sections~\ref{ch:02}--\ref{ch:05} have discussed the theoretical basis, methods, and applications of LLMs in psychology. This section examines the technical limitations and ethical challenges that must be addressed before responsible deployment.

\subsection{Technical Challenges}

\textbf{Reliability and Stability.}
Reliability, defined as the consistency of measurements across time and conditions, is a foundational requirement for psychological assessment. LLMs exhibit instability along two dimensions: prompt sensitivity and temporal inconsistency.

Prompt fragility refers to the phenomenon where minor lexical or syntactic variations in input alter model outputs substantially~\cite{cao2024worst}. Salinas et al.~\cite{salinas2024butterfly} demonstrated that semantically equivalent prompt variations can produce performance differences exceeding 20\%, linking assessment outcomes to phrasing rather than underlying constructs. In clinical contexts, practitioners' natural language variations could introduce systematic measurement error.
Temporal instability presents an equally severe challenge. Huang et al.~\cite{huang2024reliability} conducted rigorous psychometric evaluations across multiple personality instruments, revealing that LLMs generate divergent judgments for identical inputs across repeated administrations. Their findings show that even with deterministic sampling (temperature=0), test-retest reliability coefficients frequently fall below 0.70, well below the 0.90 threshold required for high-stakes clinical decisions. Bodro{\v{z}}a et al.~\cite{bodrovza2024personality} corroborated these concerns, documenting limited temporal stability in personality assessments despite models exhibiting consistent prosocial tendencies.
Wang et al.~\cite{wang2025raters} provided a comprehensive framework for assessing LLM annotation reliability. While their results demonstrate acceptable inter-rater reliability between LLM annotations and human judgments for certain tasks, they emphasize that reliability varies substantially across psychological constructs and assessment contexts. These findings underscore the necessity for construct-specific validation rather than assuming generalized reliability across all psychometric applications.

\textbf{Evaluation Challenges and Computational Costs.}
The computational demands of LLM-based psychological assessment also present barriers to widespread adoption. Traditional dictionary-based methods require minimal computational resources, typically executing in milliseconds on standard hardware through simple pattern matching and word counting. In contrast, LLM-based approaches demand orders of magnitude greater resources. Rathje et al.~\cite{rathje2024gpt} demonstrated that while GPT-based psychological text analysis achieves superior accuracy across multilingual contexts, processing equivalent text volumes requires approximately 10,000-fold increase in computational operations compared to dictionary methods, translating to significant time and energy costs.

The financial implications of commercial API deployment further constrain accessibility. At pricing structures in 2023-2024, analyzing 10,000 clinical assessment transcripts using GPT-4 (averaging 500 tokens per input) costs approximately \$375 for API calls alone, excluding infrastructure, storage, and processing overhead. For large-scale psychological research or clinical screening programs, these expenses rapidly become prohibitive. Dependency on commercial APIs also introduces privacy risks, reproducibility concerns, and vendor lock-in that conflict with research ethics and clinical standards.

Recent advances in model efficiency and open-source alternatives signal potential cost reductions. The trajectory from GPT-4's initial pricing (\$30 per million input tokens in 2023) to more affordable models like DeepSeek-R1~\cite{guo2025deepseek} (approximately \$0.14 per million tokens) represents over 200-fold cost decrease within two years. Despite these encouraging trends, LLM-based methods remain substantially more expensive than traditional approaches. A typical large-scale personality assessment study processing 100,000 questionnaire responses would incur near-zero marginal costs with dictionary methods, versus \$500--\$5,000 in API costs even with current low-cost models, excluding computational infrastructure for self-hosted alternatives. This cost disparity limits feasibility for resource-constrained research contexts and raises equity concerns regarding access to advanced assessment technologies. Balancing the enhanced capabilities of LLMs against their computational overhead remains a critical challenge for sustainable deployment in psychological practice.

\subsection{Ethical Considerations}

\textbf{Safety.}
Safety concerns encompass psychological safety (protecting users from harm during interaction) and system safety (robust failure handling). Li et al.~\cite{li2024evaluating} identified persistent vulnerabilities: models frequently generate stereotyping content, inappropriate self-harm advice, or responses that normalize harmful behaviors. Guo et al.~\cite{guo2024large} documented cases where chatbots provided inappropriate reassurances to users expressing suicidal ideation. The fundamental limitation is that LLMs lack access to behavioral indicators (e.g., tone of voice, response latency), contextual information, and causal reasoning needed for genuine risk assessment.

Therapeutic relationship disruption is another concern. Laranjo et al.~\cite{laranjo2018conversational} showed that conversational agents cannot replicate the therapeutic alliance essential for treatment. When individuals substitute LLM interactions for professional care, treatable conditions may progress.


\textbf{Privacy and Data Security.}
Psychological assessment data is highly sensitive. LLMs introduce privacy vulnerabilities because they often require transmitting raw conversational transcripts to external servers, creating exposure points for data breaches.

Kim et al.~\cite{kim2023propile} developed ProPILE, a probing tool demonstrating that personally identifiable information (PII) embedded in training data can be extracted through carefully crafted prompts, empowering data subjects to assess privacy intrusion risks. Their work on OPT-1.3B revealed that even seemingly anonymized datasets leak sensitive PII when adversaries possess auxiliary information. Hong et al.~\cite{hong2024dp} proposed DP-OPT, which employs differentially-private prompt tuning to generate privacy-preserving prompts through ensemble in-context learning, achieving competitive performance while protecting sensitive training data. Xiao et al.~\cite{xiao2024large} introduced Contextual Privacy Protection Language Models (CPPLM), demonstrating that LLMs can be fine-tuned to inject domain knowledge while safeguarding inference-time privacy through instruction-based tuning with both positive and negative examples. These approaches show promise, yet face practical limitations: differential privacy mechanisms degrade model performance, privacy budgets require careful allocation across use cases, and self-hosted deployments demand substantial infrastructure investments beyond most clinical settings' capabilities. The asymmetric information dynamics between providers and users further compound challenges, as individuals undergoing assessment often lack expertise to evaluate privacy risks, rendering truly informed consent practically unattainable.

\textbf{Clinical and Regulatory Constraints.}
The deployment of LLM-based psychological assessment tools must be situated within existing regulatory frameworks for digital mental health. The U.S.\ FDA classifies software that provides clinical decision support as a medical device when it is intended to inform diagnosis or treatment~\cite{muehlematter2021approval}. In the EU, the AI Act categorizes systems used in health-related contexts as high-risk, requiring conformity assessments and post-market surveillance. The American Psychological Association's guidelines on digital mental health emphasize that automated tools should supplement rather than replace clinical judgment, and that their psychometric properties must meet the same standards as traditional instruments~\cite{APA2013Telepsychology}. For now, no LLM-based psychological assessment tool has undergone formal regulatory clearance. These constraints underscore the gap between technical capability and permissible deployment.

\textbf{Bias and Fairness.}
Systemic biases in LLM training data pose validity challenges for psychological assessment. Personality frameworks such as the Big Five were developed and validated predominantly on Western, Educated, Industrialized, Rich, and Democratic (WEIRD) populations~\cite{santurkar2023opinions}, and LLMs absorbing English-dominant internet text inherit these cultural presuppositions~\cite{ramezani2023knowledge}. Shen et al.~\cite{shen2024cultural_bias} evaluated cultural alignment across 107 countries for five consecutive GPT versions and found persistent bias favoring Western cultural values. When deployed across culturally diverse populations, such models risk imposing culturally specific norms, undermining construct validity in cross-cultural assessment contexts.

At the clinical level, mental health datasets systematically under-represent racial and ethnic minorities, leading models to learn symptom patterns calibrated to majority-group presentations. Zack et al.~\cite{zack2024assessing} found that GPT-4 generated recommendations that differed by patient race and gender even when clinical facts were held constant. Bias also surfaces in trait attribution: linguistic markers of gender, race, or socioeconomic status can function as confounders, causing stylistic variation to influence trait inference~\cite{navigli2023biases,kotek2023gender}.  Rigorous demographic auditing and culturally diverse benchmark datasets are therefore necessary before deploying LLMs in clinical assessment pipelines.

\section{Future Directions}
\label{ch:07}

This section outlines key directions for advancing LLM-based psychological measurement toward more reliable, valid, and responsible applications.

\subsection{Technical Innovation}

The most pressing technical challenge for LLM-based psychological assessment lies in achieving reliable and reproducible measurements. As documented in Section~\ref{ch:06}, current LLMs exhibit substantial instability manifested through prompt sensitivity and temporal inconsistency, with test-retest reliability coefficients frequently falling below clinical acceptability thresholds. Addressing these limitations requires coordinated advances across multiple fronts.

Hybrid architectures combining LLM reasoning (CoT) capabilities with structured psychometric models offer another promising direction. By constraining LLM outputs to conform to psychologically validated response patterns and factor structures, such systems could provide more stable measurements while retaining the flexibility needed for nuanced interpretation. 

\subsection{Application Expansion}

Transforming research prototypes into clinical applications requires systematic mapping to real-world deployment contexts and rigorous validation. Existing LLM-based psychological assessments mainly focus on proof-of-concept demonstrations, advancing toward clinical utility demands comprehensive development of production-ready systems.

A critical direction involves developing robust downstream applications that integrate with existing clinical workflows. Such applications must provide interpretable outputs that present model assessments with explicit reasoning chains, uncertainty estimates, and relevant supporting evidence from psychological knowledge bases. Clinicians require not only diagnostic suggestions but also transparent justifications that enable them to evaluate the validity of AI recommendations and maintain independent clinical judgment. 
Specialized clinical populations demand careful validation and adaptation. Extending LLM-based assessment to children, older adults, individuals with cognitive impairments, and culturally diverse communities requires developing corresponding paradigms, and addressing linguistic and cultural factors that influence psychological expression. Before clinical deployment, such studies must examine accuracy and potential failure modes, demographic fairness, and impact on clinical decision-making and patient outcomes. The goal is not merely technological capability demonstration but rather evidence that LLM-based tools genuinely improve clinical care quality, efficiency, or accessibility without introducing unacceptable risks.
\section{Conclusion}

This survey has examined the theoretical foundations, methodological paradigms, applications, and challenges of LLM-based psychological measurement. Current evidence suggests that LLMs can support structured assessment tasks such as personality trait inference and mental health screening with moderate validity, but they fall short of the reliability and interpretability standards required for clinical decision-making. Key limitations include prompt sensitivity, temporal inconsistency, cultural bias, and the absence of regulatory approval for clinical deployment.

We emphasize that LLMs should be understood as tools to augment, not replace, traditional psychometric instruments and clinical judgment. The most promising path forward lies in hybrid systems that combine LLM-driven automation with the rigor of validated assessment frameworks. Progress will require collaboration among psychologists, computer scientists, and ethicists, guided by psychometric standards and responsible deployment principles.

\section*{Acknowledgment}

This work was supported by the National Natural Science Foundation of China under Grant 62576216 and 32441113, and Guangdong Provincial Key Laboratory under Grant 2023B1212060076

\balance

\bibliographystyle{IEEEtran}
\bibliography{references/references}

@article{amadori2023user,
  title={User-aware multilevel cognitive workload estimation from multimodal physiological signals},
  author={Amadori, Pierluigi Vito and Demiris, Yiannis},
  journal={IEEE Transactions on Cognitive and Developmental Systems},
  volume={16},
  number={4},
  pages={1212--1222},
  year={2023},
  publisher={IEEE}
}

@article{street2024llms,
  title={Llms achieve adult human performance on higher-order theory of mind tasks},
  author={Street, Winnie and Siy, John Oliver and Keeling, Geoff and Baranes, Adrien and Barnett, Benjamin and McKibben, Michael and Kanyere, Tatenda and Lentz, Alison and Dunbar, Robin IM and others},
  journal={arXiv preprint arXiv:2405.18870},
  year={2024}
}

@inproceedings{reichenpfader2024simulating,
  title={Simulating diverse patient populations using patient vignettes and large language models},
  author={Reichenpfader, Daniel and Denecke, Kerstin},
  booktitle={Proceedings of the First Workshop on Patient-Oriented Language Processing (CL4Health)@ LREC-COLING 2024},
  pages={20--25},
  year={2024}
}

@inproceedings{sorokovikova2024llms,
  title={LLMs Simulate Big Five Personality Traits: Further Evidence},
  author={Sorokovikova, Aleksandra and Fedorova, Natalia and Rezagholi, Sharwin and Wien, Technikum and Yamshchikov, Ivan P},
  booktitle={The 1st Workshop on Personalization of Generative AI Systems},
  pages={83},
  year={2024}
}

@inproceedings{moon2024virtual,
  title={Virtual Personas for Language Models via an Anthology of Backstories},
  author={Moon, Suhong and Abdulhai, Marwa and Kang, Minwoo and Suh, Joseph and Soedarmadji, Widyadewi and Behar, Eran and Chan, David},
  booktitle={Proceedings of the 2024 Conference on Empirical Methods in Natural Language Processing},
  pages={19864--19897},
  year={2024}
}

@article{ashton2007empirical,
  title={Empirical, theoretical, and practical advantages of the HEXACO model of personality structure},
  author={Ashton, Michael C and Lee, Kibeom},
  journal={Personality and social psychology review},
  volume={11},
  number={2},
  pages={150--166},
  year={2007},
  publisher={SAGE publications Sage CA: Los Angeles, CA}
}

@article{goldberg1993structure,
  title={The structure of phenotypic personality traits.},
  author={Goldberg, Lewis R},
  journal={American psychologist},
  volume={48},
  number={1},
  pages={26},
  year={1993},
  publisher={American Psychological Association}
}

@article{broska2025mixed,
  title={The Mixed Subjects Design: Treating Large Language Models as Potentially Informative Observations},
  author={Broska, David and Howes, Michael and van Loon, Austin},
  journal={Sociological Methods \& Research},
  pages={00491241251326865},
  year={2025},
  publisher={SAGE Publications Sage CA: Los Angeles, CA}
}

@article{toubia2025database,
  title={Database report: Twin-2k-500: A data set for building digital twins of over 2,000 people based on their answers to over 500 questions},
  author={Toubia, Olivier and Gui, George Z and Peng, Tianyi and Merlau, Daniel J and Li, Ang and Chen, Haozhe},
  journal={Marketing Science},
  year={2025},
  publisher={INFORMS}
}

@article{cheung2025large,
  title={Large language models show amplified cognitive biases in moral decision-making},
  author={Cheung, Vanessa and Maier, Maximilian and Lieder, Falk},
  journal={Proceedings of the National Academy of Sciences},
  volume={122},
  number={25},
  pages={e2412015122},
  year={2025},
  publisher={National Academy of Sciences}
}

@inproceedings{ramezani2023knowledge,
  title={Knowledge of cultural moral norms in large language models},
  author={Ramezani, Aida and Xu, Yang},
  booktitle={Proceedings of the 61st Annual Meeting of the Association for Computational Linguistics (Volume 1: Long Papers)},
  pages={428--446},
  year={2023}
}

@article{dong2025humanizing,
  title={Humanizing llms: A survey of psychological measurements with tools, datasets, and human-agent applications},
  author={Dong, Wenhan and Zhao, Yuemeng and Sun, Zhen and Liu, Yule and Peng, Zifan and Zheng, Jingyi and Zhang, Zongmin and Zhang, Ziyi and Wu, Jun and Wang, Ruiming and others},
  journal={arXiv preprint arXiv:2505.00049},
  year={2025}
}

@article{guo2024large,
  title={Large language models for mental health applications: systematic review},
  author={Guo, Zhijun and Lai, Alvina and Thygesen, Johan H and Farrington, Joseph and Keen, Thomas and Li, Kezhi and others},
  journal={JMIR mental health},
  volume={11},
  number={1},
  pages={e57400},
  year={2024},
  publisher={JMIR Publications Inc., Toronto, Canada}
}

@article{ke2025exploring,
  title={Exploring the frontiers of llms in psychological applications: A comprehensive review},
  author={Ke, Luoma and Tong, Song and Cheng, Peng and Peng, Kaiping},
  journal={Artificial Intelligence Review},
  volume={58},
  number={10},
  pages={305},
  year={2025},
  publisher={Springer}
}

@article{ye2025large,
  title={Large language model psychometrics: A systematic review of evaluation, validation, and enhancement},
  author={Ye, Haoran and Jin, Jing and Xie, Yuhang and Zhang, Xin and Song, Guojie},
  journal={arXiv preprint arXiv:2505.08245},
  year={2025}
}

@article{guo2025deepseek,
  title={Deepseek-r1: Incentivizing reasoning capability in llms via reinforcement learning},
  author={Guo, Daya and Yang, Dejian and Zhang, Haowei and Song, Junxiao and Zhang, Ruoyu and Xu, Runxin and Zhu, Qihao and Ma, Shirong and Wang, Peiyi and Bi, Xiao and others},
  journal={arXiv preprint arXiv:2501.12948},
  year={2025}
}

@article{achiam2023gpt,
	title={Gpt-4 technical report},
	author={Achiam, Josh and Adler, Steven and Agarwal, Sandhini and Ahmad, Lama and Akkaya, Ilge and Aleman, Florencia Leoni and Almeida, Diogo and Altenschmidt, Janko and Altman, Sam and Anadkat, Shyamal and others},
	journal={arXiv preprint arXiv:2303.08774},
	year={2023}
}

@misc{google2025gemini2,
	author       = {Shrestha Basu Mallick and Logan Kilpatrick},
	title        = {Gemini 2.0: Flash, Flash-Lite and Pro},
	year         = {2025},
	month        = {February},
	url          = {https://developers.googleblog.com/zh-hans/gemini-2-family-expands/},
	note         = {Accessed: 2025-05-01}
}

@article{reid2024gemini,
  title={Gemini 1.5: Unlocking multimodal understanding across millions of tokens of context},
  author={Reid, Machel and Savinov, Nikolay and Teplyashin, Denis and Lepikhin, Dmitry and Lillicrap, Timothy and Alayrac, Jean-Baptiste and others},
  journal={arXiv preprint arXiv:2403.05530},
  year={2024}
}

@article{wang2024qwen2vl,
  title={Qwen2-VL: Enhancing Vision-Language Model's Perception of the World at Any Resolution},
  author={Wang, Peng and Li, Xuebo and Wang, Zhijie and others},
  journal={arXiv preprint arXiv:2409.12191},
  year={2024}
}

@article{lian2023gpt4vemotion,
  title={GPT-4V with Emotion: A Zero-shot Benchmark for Multimodal Emotion Understanding},
  author={Lian, Zheng and Sun, Licai and Sun, Haiyang and Chen, Kang and Wen, Zhuofan and Gu, Hao and Chen, Shun and Liu, Bin and Tao, Jianhua},
  journal={arXiv preprint arXiv:2312.04293},
  year={2023}
}

@inproceedings{chen2023soulchat,
  title={SoulChat: Improving LLMs' Empathy, Listening, and Comfort Abilities through Fine-tuning with Multi-turn Empathy Conversations},
  author={Chen, Yirong and Xing, Xiaofen and Lin, Jingkai and Zheng, Huimin and Wang, Zhenyu and Liu, Qi and Xu, Xiangmin},
  booktitle={Findings of the Association for Computational Linguistics: EMNLP 2023},
  pages={1170--1183},
  year={2023}
}

@article{qiu2023smile,
  title={Smile: Single-turn to multi-turn inclusive language expansion via chatgpt for mental health support},
  author={Qiu, Huachuan and He, Hongliang and Zhang, Shuai and Li, Anqi and Lan, Zhenzhong},
  journal={arXiv preprint arXiv:2305.00450},
  year={2023}
}

@article{hu2024psycollm,
  title={Psycollm: Enhancing llm for psychological understanding and evaluation},
  author={Hu, Jinpeng and Dong, Tengteng and Gang, Luo and Ma, Hui and Zou, Peng and Sun, Xiao and Guo, Dan and Yang, Xun and Wang, Meng},
  journal={IEEE Transactions on Computational Social Systems},
  year={2024},
  publisher={IEEE}
}

@inproceedings{bender2021dangers,
  title={On the dangers of stochastic parrots: Can language models be too big?},
  author={Bender, Emily M and Gebru, Timnit and McMillan-Major, Angelina and Shmitchell, Shmargaret},
  booktitle={Proceedings of the 2021 ACM conference on fairness, accountability, and transparency},
  pages={610--623},
  year={2021}
}

@article{kosinski2023theory,
  title={Theory of mind may have spontaneously emerged in large language models},
  author={Kosinski, Michal},
  journal={arXiv preprint arXiv:2302.02083},
  volume={4},
  pages={169},
  year={2023}
}

@article{chen2024tombench,
  title={ToMBench: Benchmarking Theory of Mind in Large Language Models},
  author={Chen, Zhuang and Wu, Jincenzi and Zhou, Jinfeng and Wen, Bosi and Bi, Guanqun and Jiang, Gongyao and Cao, Yaru and Hu, Mengting and Lai, Yunghwei and Xiong, Zexuan and others},
  journal={CoRR},
  year={2024}
}

@article{ullman2023large,
  title={Large language models fail on trivial alterations to theory-of-mind tasks},
  author={Ullman, Tomer},
  journal={arXiv preprint arXiv:2302.08399},
  year={2023}
}

@inproceedings{wilf2024think,
  title={Think Twice: Perspective-Taking Improves Large Language Models' Theory-of-Mind Capabilities},
  author={Wilf, Alex and Lee, Sihyun and Liang, Paul Pu and Morency, Louis-Philippe},
  booktitle={Proceedings of the 62nd Annual Meeting of the Association for Computational Linguistics (Volume 1: Long Papers)},
  pages={8292--8308},
  year={2024}
}

@inproceedings{sabour2024emobench,
  title={EmoBench: Evaluating the Emotional Intelligence of Large Language Models},
  author={Sabour, Sahand and Liu, Siyang and Zhang, Zheyuan and Liu, June and Zhou, Jinfeng and Sunaryo, Alvionna and Lee, Tatia and Mihalcea, Rada and Huang, Minlie},
  booktitle={Proceedings of the 62nd Annual Meeting of the Association for Computational Linguistics (Volume 1: Long Papers)},
  pages={5986--6004},
  year={2024}
}

@article{suhr2023challenging,
  title={Challenging the validity of personality tests for large language models},
  author={S{\"u}hr, Tom and Dorner, Florian E and Samadi, Samira and Kelava, Augustin},
  journal={arXiv preprint arXiv:2311.05297},
  year={2023}
}

@inproceedings{sileo2023mindgames,
  title={MindGames: Targeting Theory of Mind in Large Language Models with Dynamic Epistemic Modal Logic},
  author={Sileo, Damien and Lernould, Antoine},
  booktitle={Findings of the Association for Computational Linguistics: EMNLP 2023},
  pages={4570--4577},
  year={2023}
}

@article{strachan2024testing,
  title={Testing theory of mind in large language models and humans},
  author={Strachan, James WA and Albergo, Dalila and Borghini, Giulia and Pansardi, Oriana and Scaliti, Eugenio and Gupta, Saurabh and Saxena, Krati and Rufo, Alessandro and Panzeri, Stefano and Manzi, Guido and others},
  journal={Nature Human Behaviour},
  volume={8},
  number={7},
  pages={1285--1295},
  year={2024},
  publisher={Nature Publishing Group UK London}
}

@article{schlinger2009theory,
  title={Theory of mind: An overview and behavioral perspective},
  author={Schlinger Jr, Henry D},
  journal={The Psychological Record},
  volume={59},
  number={3},
  pages={435--448},
  year={2009},
  publisher={Springer}
}

@inproceedings{van2023theory,
  title={Theory of Mind in Large Language Models: Examining Performance of 11 State-of-the-Art models vs. Children Aged 7-10 on Advanced Tests},
  author={van Duijn, Max and Van Dijk, Bram and Kouwenhoven, Tom and de Valk, Werner and Spruit, Marco and van der Putten, Peter},
  booktitle={Proceedings of the 27th Conference on Computational Natural Language Learning (CoNLL)},
  pages={389--402},
  year={2023}
}

@misc{kosinski2024evaluatinglargelanguagemodels,
      title={Evaluating Large Language Models in Theory of Mind Tasks}, 
      author={Michal Kosinski},
      year={2024},
      eprint={2302.02083},
      archivePrefix={arXiv},
      primaryClass={cs.CL},
      doi={https://doi.org/10.1073/pnas.2405460121},
      url={https://arxiv.org/abs/2302.02083}, 
}

@book{baron1997mindblindness,
  title={Mindblindness: An essay on autism and theory of mind},
  author={Baron-Cohen, Simon},
  year={1997},
  publisher={MIT press}
}

@inproceedings{sap2022neural,
  title={Neural Theory-of-Mind? On the Limits of Social Intelligence in Large LMs},
  author={Sap, Maarten and Le Bras, Ronan and Fried, Daniel and Choi, Yejin},
  booktitle={Proceedings of the 2022 Conference on Empirical Methods in Natural Language Processing},
  pages={3762--3780},
  year={2022}
}

@article{moghaddam2023boosting,
  title={Boosting theory-of-mind performance in large language models via prompting},
  author={Moghaddam, Shima Rahimi and Honey, Christopher J},
  journal={arXiv preprint arXiv:2304.11490},
  year={2023}
}

@article{argyle2023out,
  title={Out of one, many: Using language models to simulate human samples},
  author={Argyle, Lisa P and Busby, Ethan C and Fulda, Nancy and Gubler, Joshua R and Rytting, Christopher and Wingate, David},
  journal={Political Analysis},
  volume={31},
  number={3},
  pages={337--351},
  year={2023},
  publisher={Cambridge University Press}
}

@article{wang2025evaluating,
  title={Evaluating the ability of large language models to emulate personality},
  author={Wang, Yilei and Zhao, Jiabao and Ones, Deniz S and He, Liang and Xu, Xin},
  journal={Scientific reports},
  volume={15},
  number={1},
  pages={519},
  year={2025},
  publisher={Nature Publishing Group UK London}
}

@article{y2022large,
  title={Do large language models understand us?},
  author={y Arcas, Blaise Ag{\"u}era},
  journal={Daedalus},
  volume={151},
  number={2},
  pages={183--197},
  year={2022},
  publisher={MIT Press One Rogers Street, Cambridge, MA 02142-1209, USA journals-info~…}
}

@article{gurnee2023language,
  title={Language models represent space and time},
  author={Gurnee, Wes and Tegmark, Max},
  journal={arXiv preprint arXiv:2310.02207},
  year={2023}
}

@article{mitchell2023debate,
  title={The debate over understanding in AI's large language models},
  author={Mitchell, Melanie and Krakauer, David C},
  journal={Proceedings of the National Academy of Sciences},
  volume={120},
  number={13},
  pages={e2215907120},
  year={2023},
  publisher={National Academy of Sciences}
}

@inproceedings{bender2020climbing,
  title={Climbing towards NLU: On meaning, form, and understanding in the age of data},
  author={Bender, Emily M and Koller, Alexander},
  booktitle={Proceedings of the 58th annual meeting of the association for computational linguistics},
  pages={5185--5198},
  year={2020}
}

@article{yan2024large,
  title={Do large language models understand logic or just mimick context?},
  author={Yan, Junbing and Wang, Chengyu and Huang, Jun and Zhang, Wei},
  journal={arXiv preprint arXiv:2402.12091},
  year={2024}
}

@article{wimmer1983beliefs,
  title={Beliefs about beliefs: Representation and constraining function of wrong beliefs in young children's understanding of deception},
  author={Wimmer, Heinz and Perner, Josef},
  journal={Cognition},
  volume={13},
  number={1},
  pages={103--128},
  year={1983},
  publisher={Elsevier}
}

@article{baron1985does,
  title={Does the autistic child have a ``theory of mind''?},
  author={Baron-Cohen, Simon and Leslie, Alan M and Frith, Uta},
  journal={Cognition},
  volume={21},
  number={1},
  pages={37--46},
  year={1985},
  publisher={Elsevier}
}

@article{perner1987three,
  title={Three-year-olds' difficulty with false belief: The case for a conceptual deficit},
  author={Perner, Josef and Leekam, Susan R and Wimmer, Heinz},
  journal={British journal of developmental psychology},
  volume={5},
  number={2},
  pages={125--137},
  year={1987},
  publisher={Wiley Online Library}
}

@article{goldstein2022shared,
  title={Shared computational principles for language processing in humans and deep language models},
  author={Goldstein, Ariel and Zada, Zaid and Buchnik, Eliav and Schain, Mariano and Price, Amy and Aubrey, Bobbi and Nastase, Samuel A and Feder, Amir and Emanuel, Dotan and Cohen, Alon and others},
  journal={Nature neuroscience},
  volume={25},
  number={3},
  pages={369--380},
  year={2022},
  publisher={Nature Publishing Group US New York}
}

@inproceedings{tuna2024effects,
  title={Effects of language-and culture-specific prompting on ChatGPT},
  author={Tuna, Mustafa and Schaaff, Kristina and Schlippe, Tim},
  booktitle={2024 2nd International Conference on Foundation and Large Language Models (FLLM)},
  pages={73--81},
  year={2024},
  organization={IEEE}
}

@article{kovavc2023large,
  title={Large language models as superpositions of cultural perspectives},
  author={Kova{\v{c}}, Grgur and Sawayama, Masataka and Portelas, R{\'e}my and Colas, C{\'e}dric and Dominey, Peter Ford and Oudeyer, Pierre-Yves},
  journal={arXiv preprint arXiv:2307.07870},
  year={2023}
}

@article{wang2024patient,
  title={Patient-$\{$$\backslash$Psi$\}$: Using large language models to simulate patients for training mental health professionals},
  author={Wang, Ruiyi and Milani, Stephanie and Chiu, Jamie C and Zhi, Jiayin and Eack, Shaun M and Labrum, Travis and Murphy, Samuel M and Jones, Nev and Hardy, Kate and Shen, Hong and others},
  journal={arXiv preprint arXiv:2405.19660},
  year={2024}
}

@article{goldberg1998demographic,
  title={Demographic variables and personality: The effects of gender, age, education, and ethnic/racial status on self-descriptions of personality attributes},
  author={Goldberg, Lewis R and Sweeney, Dennis and Merenda, Peter F and Hughes Jr, John Edward},
  journal={Personality and Individual differences},
  volume={24},
  number={3},
  pages={393--403},
  year={1998},
  publisher={Elsevier}
}

@article{allport1937personality,
  title={Personality: A psychological interpretation.},
  author={Allport, Gordon Willard},
  year={1937},
  publisher={Holt}
}

@book{myers2003mbti,
  title={MBTI manual: A guide to the development and use of the Myers-Briggs Type Indicator},
  author={Myers, Isabel Briggs},
  year={2003},
  publisher={Cpp}
}

@article{lee2023chain,
  title={Chain of empathy: Enhancing empathetic response of large language models based on psychotherapy models},
  author={Lee, Yoon Kyung and Lee, Inju and Shin, Minjung and Bae, Seoyeon and Hahn, Sowon},
  journal={arXiv preprint arXiv:2311.04915},
  year={2023}
}

@inproceedings{sharma2020empathy,
    title={A Computational Approach to Understanding Empathy Expressed in Text-Based Mental Health Support},
    author={Sharma, Ashish and Miner, Adam S and Atkins, David C and Althoff, Tim},
    year={2020},
    booktitle={EMNLP}
}

@inproceedings{yang-etal-2023-psycot,
    title = "{P}sy{C}o{T}: Psychological Questionnaire as Powerful Chain-of-Thought for Personality Detection",
    author = "Yang, Tao  and
      Shi, Tianyuan  and
      Wan, Fanqi  and
      Quan, Xiaojun  and
      Wang, Qifan  and
      Wu, Bingzhe  and
      Wu, Jiaxiang",
    booktitle = "Findings of the Association for Computational Linguistics: EMNLP 2023",
    month = dec,
    year = "2023",
    address = "Singapore",
    publisher = "Association for Computational Linguistics",
    url = "https://aclanthology.org/2023.findings-emnlp.216/",
    pages = "3305--3320",
}

@article{ren2024wundtgpt,
  title={Wundtgpt: Shaping large language models to be an empathetic, proactive psychologist},
  author={Ren, Chenyu and Zhang, Yazhou and He, Daihai and Qin, Jing},
  journal={arXiv preprint arXiv:2406.15474},
  year={2024}
}

@article{hu2025beyond,
  title={Beyond Empathy: Integrating Diagnostic and Therapeutic Reasoning with Large Language Models for Mental Health Counseling},
  author={Hu, He and Zhou, Yucheng and Si, Juzheng and Wang, Qianning and Zhang, Hengheng and Ren, Fuji and Ma, Fei and Cui, Laizhong},
  journal={arXiv preprint arXiv:2505.15715},
  year={2025}
}

@article{nie2024llm,
  title={LLM-based conversational AI therapist for daily functioning screening and psychotherapeutic intervention via everyday smart devices},
  author={Nie, Jingping and Shao, Hanya and Fan, Yuang and Shao, Qijia and You, Haoxuan and Preindl, Matthias and Jiang, Xiaofan},
  journal={arXiv preprint arXiv:2403.10779},
  year={2024}
}

@inproceedings{qiu2024psychat,
  title={Psychat: A client-centric dialogue system for mental health support},
  author={Qiu, Huachuan and Li, Anqi and Ma, Lizhi and Lan, Zhenzhong},
  booktitle={2024 27th International Conference on Computer Supported Cooperative Work in Design (CSCWD)},
  pages={2979--2984},
  year={2024},
  organization={IEEE}
}

@article{rathje2024gpt,
  title={GPT is an effective tool for multilingual psychological text analysis},
  author={Rathje, Steve and Mirea, Dan-Mircea and Sucholutsky, Ilia and Marjieh, Raja and Robertson, Claire E and Van Bavel, Jay J},
  journal={Proceedings of the National Academy of Sciences},
  volume={121},
  number={34},
  pages={e2308950121},
  year={2024},
  publisher={National Academy of Sciences}
}

@article{niu2024text,
  title={From text to emotion: Unveiling the emotion annotation capabilities of llms},
  author={Niu, Minxue and Jaiswal, Mimansa and Provost, Emily Mower},
  journal={arXiv preprint arXiv:2408.17026},
  year={2024}
}

@article{rao2023can,
  title={Can chatgpt assess human personalities? a general evaluation framework},
  author={Rao, Haocong and Leung, Cyril and Miao, Chunyan},
  journal={arXiv preprint arXiv:2303.01248},
  year={2023}
}

@inproceedings{hu2024llm,
  title={Llm vs small model? large language model based text augmentation enhanced personality detection model},
  author={Hu, Linmei and He, Hongyu and Wang, Duokang and Zhao, Ziwang and Shao, Yingxia and Nie, Liqiang},
  booktitle={Proceedings of the AAAI Conference on Artificial Intelligence},
  volume={38},
  number={16},
  pages={18234--18242},
  year={2024}
}

@article{teng2025enhancing,
  title={Enhancing Depression Detection with Chain-of-Thought Prompting: From Emotion to Reasoning Using Large Language Models},
  author={Teng, Shiyu and Liu, Jiaqing and Jain, Rahul Kumar and Chai, Shurong and Hou, Ruibo and Tateyama, Tomoko and Lin, Lanfen and Chen, Yen-wei},
  journal={arXiv preprint arXiv:2502.05879},
  year={2025}
}

@inproceedings{ringeval2019avec,
  title={AVEC 2019 workshop and challenge: state-of-mind, detecting depression with AI, and cross-cultural affect recognition},
  author={Ringeval, Fabien and Schuller, Bj{\"o}rn and Valstar, Michel and Cummins, Nicholas and Cowie, Roddy and Tavabi, Leili and Schmitt, Maximilian and Alisamir, Sina and Amiriparian, Shahin and Messner, Eva-Maria and others},
  booktitle={Proceedings of the 9th International on Audio/visual Emotion Challenge and Workshop},
  pages={3--12},
  year={2019}
}

@article{xu2024mental,
  title={Mental-llm: Leveraging large language models for mental health prediction via online text data},
  author={Xu, Xuhai and Yao, Bingsheng and Dong, Yuanzhe and Gabriel, Saadia and Yu, Hong and Hendler, James and Ghassemi, Marzyeh and Dey, Anind K and Wang, Dakuo},
  journal={Proceedings of the ACM on Interactive, Mobile, Wearable and Ubiquitous Technologies},
  volume={8},
  number={1},
  pages={1--32},
  year={2024},
  publisher={ACM New York, NY, USA}
}

@article{shah2025advancing,
  title={Advancing depression detection on social media platforms through fine-tuned large language models},
  author={Shah, Shahid Munir and Gillani, Syeda Anshrah and Baig, Mirza Samad Ahmed and Saleem, Muhammad Aamer and Siddiqui, Muhammad Hamzah},
  journal={Online Social Networks and Media},
  volume={46},
  pages={100311},
  year={2025},
  publisher={Elsevier}
}

@inproceedings{feng2024autonomous,
  title={Autonomous aspect-image instruction a2ii: Q-former guided multimodal sentiment classification},
  author={Feng, Junjia and Lin, Mingqian and Shang, Lin and Gao, Xiaoying},
  booktitle={Proceedings of the 2024 joint international conference on computational linguistics, language resources and evaluation (LREC-COLING 2024)},
  pages={1996--2005},
  year={2024}
}

@article{yang2024empirical,
  title={An empirical study of Multimodal Entity-Based Sentiment Analysis with ChatGPT: Improving in-context learning via entity-aware contrastive learning},
  author={Yang, Li and Wang, Zengzhi and Li, Ziyan and Na, Jin-Cheon and Yu, Jianfei},
  journal={Information Processing \& Management},
  volume={61},
  number={4},
  pages={103724},
  year={2024},
  publisher={Elsevier}
}

@article{englhardt2024classification,
  title={From classification to clinical insights: Towards analyzing and reasoning about mobile and behavioral health data with large language models},
  author={Englhardt, Zachary and Ma, Chengqian and Morris, Margaret E and Chang, Chun-Cheng and Xu, Xuhai" Orson" and Qin, Lianhui and McDuff, Daniel and Liu, Xin and Patel, Shwetak and Iyer, Vikram},
  journal={Proceedings of the ACM on Interactive, Mobile, Wearable and Ubiquitous Technologies},
  volume={8},
  number={2},
  pages={1--25},
  year={2024},
  publisher={ACM New York, NY, USA}
}

@article{zheng2025promind,
  title={ProMind-LLM: Proactive Mental Health Care via Causal Reasoning with Sensor Data},
  author={Zheng, Xinzhe and Ji, Sijie and Sun, Jiawei and Chen, Renqi and Gao, Wei and Srivastava, Mani},
  journal={arXiv preprint arXiv:2505.14038},
  year={2025}
}

@article{miller2001inpatient,
  title={Inpatient diagnostic assessments: 1. Accuracy of structured vs. unstructured interviews},
  author={Miller, Paul R and Dasher, Robert and Collins, Rodney and Griffiths, Pamela and Brown, Fred},
  journal={Psychiatry research},
  volume={105},
  number={3},
  pages={255--264},
  year={2001},
  publisher={Elsevier}
}

@article{huang2025survey,
  title={A survey on hallucination in large language models: Principles, taxonomy, challenges, and open questions},
  author={Huang, Lei and Yu, Weijiang and Ma, Weitao and Zhong, Weihong and Feng, Zhangyin and Wang, Haotian and Chen, Qianglong and Peng, Weihua and Feng, Xiaocheng and Qin, Bing and others},
  journal={ACM Transactions on Information Systems},
  volume={43},
  number={2},
  pages={1--55},
  year={2025},
  publisher={ACM New York, NY}
}

@article{gilbert2015use,
  title={Use of patient-reported outcomes to measure symptoms and health related quality of life in the clinic},
  author={Gilbert, Alexandra and Sebag-Montefiore, David and Davidson, Susan and Velikova, Galina},
  journal={Gynecologic oncology},
  volume={136},
  number={3},
  pages={429--439},
  year={2015},
  publisher={Elsevier}
}

@article{brinkmann201414,
  title={14 Unstructured and Semi-Structured Interviewing},
  author={Brinkmann, Svend},
  journal={The Oxford handbook of qualitative research},
  pages={277},
  year={2014},
  publisher={Oxford University Press}
}

@article{kermani2025systematic,
  title={A Systematic Evaluation of LLM Strategies for Mental Health Text Analysis: Fine-tuning vs. Prompt Engineering vs. RAG},
  author={Kermani, Arshia and Perez-Rosas, Veronica and Metsis, Vangelis},
  journal={arXiv preprint arXiv:2503.24307},
  year={2025}
}

@article{peters2024large,
  title={Large language models can infer personality from free-form user interactions},
  author={Peters, Heinrich and Cerf, Moran and Matz, Sandra C},
  journal={arXiv preprint arXiv:2405.13052},
  year={2024}
}

@article{zhu2025investigating,
  title={Investigating Large Language Models in Inferring Personality Traits from User Conversations},
  author={Zhu, Jianfeng and Jin, Ruoming and Coifman, Karin G},
  journal={arXiv preprint arXiv:2501.07532},
  year={2025}
}

@inproceedings{wen2024affective,
  title={Affective-nli: Towards accurate and interpretable personality recognition in conversation},
  author={Wen, Zhiyuan and Cao, Jiannong and Yang, Yu and Yang, Ruosong and Liu, Shuaiqi},
  booktitle={2024 IEEE International Conference on Pervasive Computing and Communications (PerCom)},
  pages={184--193},
  year={2024},
  organization={IEEE}
}

@article{zhou2024chinese,
  title={A Chinese Multi-label Affective Computing Dataset Based on Social Media Network Users},
  author={Zhou, Jingyi and Luo, Senlin and Chen, Haofan},
  journal={arXiv preprint arXiv:2411.08347},
  year={2024}
}

@article{cao2025multimodal,
  title={A multimodal depression consultation dataset of speech and text with hamd-17 assessments},
  author={Cao, Pengfei and Zhang, Yuanzhe and Zhang, Chenxiang and Chen, Wei and Liu, Yan and Xu, Shuang and Xu, Miao and Jin, Wenqing and Xu, Jinjie and Wang, Dan and others},
  journal={Scientific Data},
  volume={12},
  number={1},
  pages={1577},
  year={2025},
  publisher={Nature Publishing Group UK London}
}

@inproceedings{qin2025mental,
  title={Mental-Perceiver: Audio-Textual Multi-Modal Learning for Estimating Mental Disorders},
  author={Qin, Jinghui and Liu, Changsong and Tang, Tianchi and Liu, Dahuang and Wang, Minghao and Huang, Qianying and Zhang, Rumin},
  booktitle={Proceedings of the AAAI Conference on Artificial Intelligence},
  volume={39},
  number={23},
  pages={25029--25037},
  year={2025}
}

@article{liu2023chatcounselor,
  title={Chatcounselor: A large language models for mental health support},
  author={Liu, June M and Li, Donghao and Cao, He and Ren, Tianhe and Liao, Zeyi and Wu, Jiamin},
  journal={arXiv preprint arXiv:2309.15461},
  year={2023}
}

@inproceedings{lee2024cactus,
  title={Cactus: Towards psychological counseling conversations using cognitive behavioral theory},
  author={Lee, Suyeon and Mac Kim, Sunghwan and Kim, Minju and Kang, Dongjin and Yang, Dongil and Kim, Harim and Kang, Minseok and Jung, Dayi and Kim, Min Hee and Lee, Seungbeen and others},
  booktitle={Findings of the Association for Computational Linguistics: EMNLP 2024},
  pages={14245--14274},
  year={2024}
}

@inproceedings{bhandari2025can,
  title={Can LLM Agents Maintain a Persona in Discourse?},
  author={Bhandari, Pranav and Fay, Nicolas and Wise, Michael J. and Datta, Amitava and Meek, Stephanie and Naseem, Usman and Nasim, Mehwish},
  booktitle={Proceedings of the 2025 Conference on Empirical Methods in Natural Language Processing},
  pages={29213--29229},
  year={2025},
  publisher={Association for Computational Linguistics},
  doi={10.18653/v1/2025.emnlp-main.1508},
  url={https://aclanthology.org/2025.emnlp-main.1508/}
}

@article{de2025introducing,
  title={Introducing CounseLLMe: A dataset of simulated mental health dialogues for comparing LLMs like Haiku, LLaMAntino and ChatGPT against humans},
  author={De Duro, Edoardo Sebastiano and Improta, Riccardo and Stella, Massimo},
  journal={Emerging Trends in Drugs, Addictions, and Health},
  volume={5},
  pages={100170},
  year={2025},
  publisher={Elsevier}
}

@article{Abbasi2024PsychoLex,
  title={PsychoLex: Unveiling the Psychological Mind of Large Language Models},
  author={Mohammad Amin Abbasi and Farnaz Sadat Mirnezami and Hassan Naderi},
}

@article{de2024phdgpt,
  title={Phdgpt: Introducing a psychometric and linguistic dataset about how large language models perceive graduate students and professors in psychology},
  author={De Duro, Edoardo Sebastiano and Taietta, Enrique and Improta, Riccardo and Stella, Massimo},
  journal={arXiv preprint arXiv:2411.10473},
  year={2024}
}

@inproceedings{mohammad2024worrywords,
  title={WorryWords: Norms of anxiety association for over 44k English words},
  author={Mohammad, Saif},
  booktitle={Proceedings of the 2024 Conference on Empirical Methods in Natural Language Processing},
  pages={16261--16278},
  year={2024}
}

@article{kosinski2013private,
  title={Private traits and attributes are predictable from digital records of human behavior},
  author={Kosinski, Michal and Stillwell, David and Graepel, Thore},
  journal={Proceedings of the national academy of sciences},
  volume={110},
  number={15},
  pages={5802--5805},
  year={2013},
  publisher={National Academy of Sciences}
}

@inproceedings{shen2022automatic,
  title={Automatic depression detection: An emotional audio-textual corpus and a gru/bilstm-based model},
  author={Shen, Ying and Yang, Huiyu and Lin, Lin},
  booktitle={ICASSP 2022-2022 IEEE International Conference on Acoustics, Speech and Signal Processing (ICASSP)},
  pages={6247--6251},
  year={2022},
  organization={IEEE}
}

@article{maharjan2025psychometric,
  title={Psychometric evaluation of large language model embeddings for personality trait prediction},
  author={Maharjan, Julina and Jin, Ruoming and Zhu, Jianfeng and Kenne, Deric},
  journal={Journal of Medical Internet Research},
  volume={27},
  pages={e75347},
  year={2025},
  publisher={JMIR Publications Toronto, Canada}
}

@article{park2023thinking,
  title={Thinking assistants: Llm-based conversational assistants that help users think by asking rather than answering},
  author={Park, Soya and Kulkarni, Chinmay},
  journal={arXiv preprint arXiv:2312.06024},
  year={2023}
}

@inproceedings{weijers-etal-2024-quantifying,
  title = {Quantifying learning-style adaptation in effectiveness of {LLM} teaching},
  author = {Weijers, Ruben and Fidelis de Castilho, Gabrielle and Godbout, Jean-Fran{\c{c}}ois and Rabbany, Reihaneh and Pelrine, Kellin},
  booktitle = {Proceedings of the 1st Workshop on Personalization of Generative AI Systems ({PERSONALIZE} 2024)},
  pages = {112--118},
  year = {2024},
  address = {St. Julians, Malta},
  publisher = {Association for Computational Linguistics},
  doi = {10.18653/v1/2024.personalize-1.10}
}

@inproceedings{ICLR2025_c9867d5a,
 author = {Li, Belinda and Tamkin, Alex and Goodman, Noah and Andreas, Jacob},
 booktitle = {International Conference on Representation Learning},
 editor = {Y. Yue and A. Garg and N. Peng and F. Sha and R. Yu},
 pages = {80984--81013},
 title = {Eliciting Human Preferences with Language Models},
 url = {https://proceedings.iclr.cc/paper_files/paper/2025/file/c9867d5a22653ce98b02595061e40f12-Paper-Conference.pdf},
 volume = {2025},
 year = {2025}
}

@inproceedings{zhang2025modeling,
title={Modeling Future Conversation Turns to Teach {LLM}s to Ask Clarifying Questions},
author={Michael JQ Zhang and W. Bradley Knox and Eunsol Choi},
booktitle={The Thirteenth International Conference on Learning Representations},
year={2025},
url={https://openreview.net/forum?id=cwuSAR7EKd}
}

@article{ma2025post,
  title={From Post To Personality: Harnessing LLMs for MBTI Prediction in Social Media},
  author={Ma, Tian and Feng, Kaiyu and Rong, Yu and Zhao, Kangfei},
  journal={arXiv preprint arXiv:2509.04461},
  year={2025}
}

@article{pennebaker1999linguistic,
  title={Linguistic styles: language use as an individual difference.},
  author={Pennebaker, James W and King, Laura A},
  journal={Journal of personality and social psychology},
  volume={77},
  number={6},
  pages={1296},
  year={1999},
  publisher={American Psychological Association}
}

@inproceedings{sun-etal-2021-psyqa,
    title = "PsyQA: A Chinese Dataset for Generating Long Counseling Text for Mental Health Support",
    author = "Sun, Hao  and
      Lin, Zhenru  and
      Zheng, Chujie  and
      Liu, Siyang  and
      Huang, Minlie",
    booktitle = "Findings of the Association for Computational Linguistics: ACL 2021",
    year = "2021",
}

@article{chen2022cped,
  title={Cped: A large-scale chinese personalized and emotional dialogue dataset for conversational ai},
  author={Chen, Yirong and Fan, Weiquan and Xing, Xiaofen and Pang, Jianxin and Huang, Minlie and Han, Wenjing and Tie, Qianfeng and Xu, Xiangmin},
  journal={arXiv preprint arXiv:2205.14727},
  year={2022}
}

@inproceedings{min2020ambigqa,
    title={ {A}mbig{QA}: Answering Ambiguous Open-domain Questions },
    author={ Min, Sewon and Michael, Julian and Hajishirzi, Hannaneh and Zettlemoyer, Luke },
    booktitle={ EMNLP },
    year={ 2020 }
}

@inproceedings{buechel-hahn-2017-emobank,
    title = "{E}mo{B}ank: Studying the Impact of Annotation Perspective and Representation Format on Dimensional Emotion Analysis",
    author = "Buechel, Sven  and
      Hahn, Udo",
    editor = "Lapata, Mirella  and
      Blunsom, Phil  and
      Koller, Alexander",
    booktitle = "Proceedings of the 15th Conference of the {E}uropean Chapter of the Association for Computational Linguistics: Volume 2, Short Papers",
    month = apr,
    year = "2017",
    address = "Valencia, Spain",
    publisher = "Association for Computational Linguistics",
    url = "https://aclanthology.org/E17-2092/",
    pages = "578--585",
}

@inproceedings{gjurkovic-etal-2021-pandora,
    title = "{PANDORA} Talks: Personality and Demographics on {R}eddit",
    author = "Gjurkovi{\'c}, Matej  and
      Karan, Vanja Mladen  and
      Vukojevi{\'c}, Iva  and
      Bo{\v{s}}njak, Mihaela  and
      Snajder, Jan",
    editor = "Ku, Lun-Wei  and
      Li, Cheng-Te",
    booktitle = "Proceedings of the Ninth International Workshop on Natural Language Processing for Social Media",
    month = jun,
    year = "2021",
    address = "Online",
    publisher = "Association for Computational Linguistics",
    url = "https://aclanthology.org/2021.socialnlp-1.12/",
    doi = "10.18653/v1/2021.socialnlp-1.12",
    pages = "138--152"
}

@misc{kaggle:mbti-type,
  title        = {{MBTI} {T}ype {D}ataset},
  author       = {{datasnaek}},
  year         = {2017},
  publisher    = {Kaggle},
  howpublished = {\url{https://www.kaggle.com/datasets/datasnaek/mbti-type/data}},
  note         = {Accessed: 2025-10-15}
}

@inproceedings{ijcai2017p536,
  author    = {Guangyao Shen and Jia Jia and Liqiang Nie and Fuli Feng and Cunjun Zhang and Tianrui Hu and Tat-Seng Chua and Wenwu Zhu},
  title     = {Depression Detection via Harvesting Social Media: A Multimodal Dictionary Learning Solution},
  booktitle = {Proceedings of the Twenty-Sixth International Joint Conference on
               Artificial Intelligence, {IJCAI-17}},
  pages     = {3838--3844},
  year      = {2017},
  doi       = {10.24963/ijcai.2017/536},
  url       = {https://doi.org/10.24963/ijcai.2017/536},
}

@article{turcan2019dreaddit,
  title={Dreaddit: A reddit dataset for stress analysis in social media},
  author={Turcan, Elsbeth and McKeown, Kathleen},
  journal={arXiv preprint arXiv:1911.00133},
  year={2019}
}

@article{zhang2018adaptive, 
    title={Adaptive Co-attention Network for Named Entity Recognition in Tweets}, 
    author={Zhang, Qi and Fu, Jinlan and Liu, Xiaoyu and Huang, Xuanjing}, 
    year={2018},
    publisher={AAAI} 
}

@inproceedings{lu2018visual,
  title={Visual attention model for name tagging in multimodal social media},
  author={Lu, Di and Neves, Leonardo and Carvalho, Vitor and Zhang, Ning and Ji, Heng},
  booktitle={Proceedings of the 56th Annual Meeting of the Association for Computational Linguistics (Volume 1: Long Papers)},
  pages={1990--1999},
  year={2018}
}

@inproceedings{
    xu2022globem_neurips,
    title={{GLOBEM} Dataset: Multi-Year Datasets for Longitudinal Human Behavior Modeling Generalization},
    author={Xuhai Xu and Han Zhang and Yasaman S Sefidgar and Yiyi Ren and Xin Liu and Woosuk Seo and Jennifer Brown and Kevin Scott Kuehn and Mike A Merrill and Paula S Nurius and Shwetak Patel and Tim Althoff and Margaret E Morris and Eve A. Riskin and Jennifer Mankoff and Anind Dey},
    booktitle={Thirty-sixth Conference on Neural Information Processing Systems Datasets and Benchmarks Track},
    year={2022},
    url={https://arxiv.org/abs/2211.02733}
}

@inproceedings{thambawita2020pmdata,
  title={Pmdata: a sports logging dataset},
  author={Thambawita, Vajira and Hicks, Steven Alexander and Borgli, Hanna and Stensland, H{\aa}kon Kvale and Jha, Debesh and Svensen, Martin Kristoffer and Pettersen, Svein-Arne and Johansen, Dag and Johansen, H{\aa}vard Dagenborg and Pettersen, Susann Dahl and others},
  booktitle={Proceedings of the 11th ACM Multimedia Systems Conference},
  pages={231--236},
  year={2020}
}

@article{colizzi2020prevention,
  title={Prevention and early intervention in youth mental health: is it time for a multidisciplinary and trans-diagnostic model for care?},
  author={Colizzi, Marco and Lasalvia, Antonio and Ruggeri, Mirella},
  journal={International journal of mental health systems},
  volume={14},
  number={1},
  pages={23},
  year={2020},
  publisher={Springer}
}

@inproceedings{verma2023ai,
  title={AI-enhanced mental health diagnosis: leveraging transformers for early detection of depression tendency in textual data},
  author={Verma, Srishti and Joshi, Rakesh Chandra and Dutta, Malay Kishore and Jezek, Stepan and Burget, Radim and others},
  booktitle={2023 15th International Congress on Ultra Modern Telecommunications and Control Systems and Workshops (ICUMT)},
  pages={56--61},
  year={2023},
  organization={IEEE}
}

@inproceedings{danner2023advancing,
  title={Advancing mental health diagnostics: GPT-based method for depression detection},
  author={Danner, Michael and Hadzic, Bakir and Gerhardt, Sophie and Ludwig, Simon and Uslu, Irem and Shao, Peng and Weber, Thomas and Shiban, Youssef and Ratsch, Matthias},
  booktitle={2023 62nd Annual Conference of the Society of Instrument and Control Engineers (SICE)},
  pages={1290--1296},
  year={2023},
  organization={IEEE}
}

@inproceedings{tao2023classifying,
  title={Classifying anxiety and depression through LLMs virtual interactions: a case study with ChatGPT},
  author={Tao, Yongfeng and Yang, Minqiang and Shen, Hao and Yang, Zhichao and Weng, Ziru and Hu, Bin},
  booktitle={2023 IEEE international conference on bioinformatics and biomedicine (BIBM)},
  pages={2259--2264},
  year={2023},
  organization={IEEE}
}

@inproceedings{sadeghi2023exploring,
  title={Exploring the capabilities of a language model-only approach for depression detection in text data},
  author={Sadeghi, Misha and Egger, Bernhard and Agahi, Reza and Richer, Robert and Capito, Klara and Rupp, Lydia Helene and Schindler-Gmelch, Lena and Berking, Matthias and Eskofier, Bjoern M},
  booktitle={2023 IEEE EMBS International Conference on Biomedical and Health Informatics (BHI)},
  pages={1--5},
  year={2023},
  organization={IEEE}
}

@inproceedings{stigall2024large,
  title={Large language models performance comparison of emotion and sentiment classification},
  author={Stigall, William and Al Hafiz Khan, Md Abdullah and Attota, Dinesh and Nweke, Francis and Pei, Yong},
  booktitle={Proceedings of the 2024 ACM Southeast Conference},
  pages={60--68},
  year={2024}
}

@article{lossio2024comparison,
  title={A comparison of ChatGPT and fine-tuned open pre-trained transformers (OPT) against widely used sentiment analysis tools: sentiment analysis of COVID-19 survey data},
  author={Lossio-Ventura, Juan Antonio and Weger, Rachel and Lee, Angela Y and Guinee, Emily P and Chung, Joyce and Atlas, Lauren and Linos, Eleni and Pereira, Francisco},
  journal={JMIR Mental Health},
  volume={11},
  pages={e50150},
  year={2024},
  publisher={JMIR Publications Toronto, Canada}
}

@article{nandwani2021review,
  title={A review on sentiment analysis and emotion detection from text},
  author={Nandwani, Pansy and Verma, Rupali},
  journal={Social network analysis and mining},
  volume={11},
  number={1},
  pages={81},
  year={2021},
  publisher={Springer}
}

@inproceedings{liu2024emollms,
  title={Emollms: A series of emotional large language models and annotation tools for comprehensive affective analysis},
  author={Liu, Zhiwei and Yang, Kailai and Xie, Qianqian and Zhang, Tianlin and Ananiadou, Sophia},
  booktitle={Proceedings of the 30th ACM SIGKDD Conference on Knowledge Discovery and Data Mining},
  pages={5487--5496},
  year={2024}
}

@article{li2025emoverse,
  title={EmoVerse: Enhancing Multimodal Large Language Models for Affective Computing via Multitask Learning},
  author={Li, Ao and Xu, Longwei and Ling, Chen and Zhang, Jinghui and Wang, Pengwei},
  journal={Neurocomputing},
  volume={650},
  pages={130810},
  year={2025},
  publisher={Elsevier}
}

@article{deng2025evaluating,
  title={Evaluating Large Language Models in Crisis Detection: A Real-World Benchmark from Psychological Support Hotlines},
  author={Deng, Guifeng and Rao, Shuyin and Lin, Tianyu and Dai, Anlu and Wang, Pan and Xie, Junyi and Song, Haidong and Zhao, Ke and Xu, Dongwu and Cheng, Zhengdong and others},
  journal={arXiv preprint arXiv:2506.01329},
  year={2025}
}

@inproceedings{wu2025psychological,
  title={Psychological health knowledge-enhanced LLM-based social network crisis intervention text transfer recognition method},
  author={Wu, Shurui and Huang, Xinyi and Lu, Dingxin},
  booktitle={Proceedings of the 2025 International Conference on Health Big Data},
  pages={156--161},
  year={2025}
}

@inproceedings{xu2024utilizing,
  title={Utilizing Large Language Models for Psychological Assessment: Enhancing Suicide Risk Detection Through Social Media Analysis},
  author={Xu, Zeling and Xu, Jieping and Luo, Yishi and Zhang, Keyu and Zhang, Jinyuan and Zou, Yuntao and Liu, Li},
  booktitle={2024 6th International Conference on Frontier Technologies of Information and Computer (ICFTIC)},
  pages={1418--1421},
  year={2024},
  organization={IEEE}
}

@article{lamsal2024crisistransformers,
  title={CrisisTransformers: Pre-trained language models and sentence encoders for crisis-related social media texts},
  author={Lamsal, Rabindra and Read, Maria Rodriguez and Karunasekera, Shanika},
  journal={Knowledge-Based Systems},
  volume={296},
  pages={111916},
  year={2024},
  publisher={Elsevier}
}

@inproceedings{diniz2022boamente,
  title={Boamente: a natural language processing-based digital phenotyping tool for smart monitoring of suicidal ideation},
  author={Diniz, Evandro JS and Fontenele, Jos{\'e} E and de Oliveira, Adonias C and Bastos, Victor H and Teixeira, Silmar and Rab{\^e}lo, Ricardo L and Cal{\c{c}}ada, Dario B and Dos Santos, Renato M and de Oliveira, Ana K and Teles, Ariel S},
  booktitle={Healthcare},
  volume={10},
  number={4},
  pages={698},
  year={2022},
  organization={MDPI}
}

@inproceedings{devlin2019bert,
  title={Bert: Pre-training of deep bidirectional transformers for language understanding},
  author={Devlin, Jacob and Chang, Ming-Wei and Lee, Kenton and Toutanova, Kristina},
  booktitle={Proceedings of the 2019 conference of the North American chapter of the association for computational linguistics: human language technologies, volume 1 (long and short papers)},
  pages={4171--4186},
  year={2019}
}

@article{yenduri2023generative,
  title={Generative pre-trained transformer: A comprehensive review on enabling technologies, potential applications, emerging challenges, and future directions},
  author={Yenduri, Gokul and Srivastava, Gautam and Maddikunta, Praveen Kumar Reddy and Jhaveri, Rutvij H and Wang, Weizheng and Vasilakos, Athanasios V and Gadekallu, Thippa Reddy and others},
  journal={arXiv preprint arXiv:2305.10435},
  year={2023}
}

@inproceedings{radford2023robust,
  title={Robust speech recognition via large-scale weak supervision},
  author={Radford, Alec and Kim, Jong Wook and Xu, Tao and Brockman, Greg and McLeavey, Christine and Sutskever, Ilya},
  booktitle={International conference on machine learning},
  pages={28492--28518},
  year={2023},
  organization={PMLR}
}

@article{liu2019roberta,
  title={Roberta: A robustly optimized bert pretraining approach},
  author={Liu, Yinhan and Ott, Myle and Goyal, Naman and Du, Jingfei and Joshi, Mandar and Chen, Danqi and Levy, Omer and Lewis, Mike and Zettlemoyer, Luke and Stoyanov, Veselin},
  journal={arXiv preprint arXiv:1907.11692},
  year={2019}
}

@article{ghanadian2024socially,
  title={Socially aware synthetic data generation for suicidal ideation detection using large language models},
  author={Ghanadian, Hamideh and Nejadgholi, Isar and Al Osman, Hussein},
  journal={IEEe Access},
  volume={12},
  pages={14350--14363},
  year={2024},
  publisher={IEEE}
}

@article{kuzmin2024mental,
  title={Mental disorders detection in the era of large language models},
  author={Kuzmin, Gleb and Strepetov, Petr and Stankevich, Maksim and Shelmanov, Artem and Smirnov, Ivan},
  journal={arXiv preprint arXiv:2410.07129},
  year={2024}
}

@article{gao2023retrieval,
  title={Retrieval-augmented generation for large language models: A survey},
  author={Gao, Yunfan and Xiong, Yun and Gao, Xinyu and Jia, Kangxiang and Pan, Jinliu and Bi, Yuxi and Sun, Jiawei and Wang, Haofen},
  journal={arXiv preprint arXiv:2312.10997},
  year={2023}
}

@article{wang2025multi,
  title={A Multi-Stage Large Language Model Framework for Extracting Suicide-Related Social Determinants of Health},
  author={Wang, Song and Wei, Yishu and Ma, Haotian and Lovitt, Max and Deng, Kelly and Meng, Yuan and Xu, Zihan and Zhang, Jingze and Xiao, Yunyu and Ding, Ying and others},
  journal={arXiv preprint arXiv:2508.05003},
  year={2025}
}

@article{gao2025leveraging,
  title={Leveraging Large Language Models for Spontaneous Speech-Based Suicide Risk Detection},
  author={Gao, Yifan and Fu, Jiao and Guo, Long and Liu, Hong},
  journal={arXiv preprint arXiv:2507.00693},
  year={2025}
}

@inproceedings{dutta2025llm,
  title={Llm supervised pre-training for multimodal emotion recognition in conversations},
  author={Dutta, Soumya and Ganapathy, Sriram},
  booktitle={ICASSP 2025-2025 IEEE International Conference on Acoustics, Speech and Signal Processing (ICASSP)},
  pages={1--5},
  year={2025},
  organization={IEEE}
}

@inproceedings{hong2025aer,
  title={AER-LLM: Ambiguity-aware emotion recognition leveraging large language models},
  author={Hong, Xin and Gong, Yuan and Sethu, Vidhyasaharan and Dang, Ting},
  booktitle={ICASSP 2025-2025 IEEE International Conference on Acoustics, Speech and Signal Processing (ICASSP)},
  pages={1--5},
  year={2025},
  organization={IEEE}
}

@article{park2015automatic,
  title={Automatic personality assessment through social media language.},
  author={Park, Gregory and Schwartz, H Andrew and Eichstaedt, Johannes C and Kern, Margaret L and Kosinski, Michal and Stillwell, David J and Ungar, Lyle H and Seligman, Martin EP},
  journal={Journal of personality and social psychology},
  volume={108},
  number={6},
  pages={934},
  year={2015},
  publisher={American Psychological Association}
}

@article{qin2025explainable,
  title={Explainable and Interactive LLMs-Augmented Depression Detection in Social Media},
  author={Qin, Wei and Chen, Zetong and Yang, Xun and Wang, Lei and Lan, Yunshi and Ren, Weijieying and Hong, Richang},
  journal={IEEE Transactions on Computational Social Systems},
  year={2025},
  publisher={IEEE}
}

@inproceedings{li2025revise,
  title={Revise, reason, and recognize: Llm-based emotion recognition via emotion-specific prompts and asr error correction},
  author={Li, Yuanchao and Gong, Yuan and Yang, Chao-Han Huck and Bell, Peter and Lai, Catherine},
  booktitle={ICASSP 2025-2025 IEEE International Conference on Acoustics, Speech and Signal Processing (ICASSP)},
  pages={1--5},
  year={2025},
  organization={IEEE}
}

@inproceedings{yang2024psychogat,
  title={PsychoGAT: A Novel Psychological Measurement Paradigm through Interactive Fiction Games with LLM Agents},
  author={Yang, Qisen and Wang, Zekun and Chen, Honghui and Wang, Shenzhi and Pu, Yifan and Gao, Xin and Huang, Wenhao and Song, Shiji and Huang, Gao},
  booktitle={Proceedings of the 62nd Annual Meeting of the Association for Computational Linguistics (Volume 1: Long Papers)},
  pages={14470--14505},
  year={2024}
}

@article{pramod2018identifying,
  title={Identifying Personality Traits using Social Media},
  author={PRAMOD, MAKKENA and Kumar, M Raj and KUMAR, PAMULA ANIL and SARATH, NIZAMPATNAM NAGA and Vikas, K},
  journal={Iconic Research and Engineering Journals},
  volume={1},
  number={9},
  pages={186--192},
  year={2018}
}

@inproceedings{tong2024advirds,
  title={ADViRDS: Assessment of Domestic Violence Risk Dataset and Scale on Social Media},
  author={Tong, Chengwei and Guo, Mengzhuo and Tian, Yao and Zhang, Mengzhu and Li, Yangyang and Zhu, Chunyan and Bao, Jie and Sheng, Rongrong and Li, Qianqian and Liao, Yong},
  booktitle={Proceedings of the Annual Meeting of the Cognitive Science Society},
  volume={46},
  year={2024}
}

@inproceedings{nielsen2022mumin,
  title={Mumin: A large-scale multilingual multimodal fact-checked misinformation social network dataset},
  author={Nielsen, Dan S and McConville, Ryan},
  booktitle={Proceedings of the 45th international ACM SIGIR conference on research and development in information retrieval},
  pages={3141--3153},
  year={2022}
}

@article{ji2023chatgpt,
  title={Is chatgpt a good personality recognizer? a preliminary study},
  author={Ji, Yu and Wu, Wen and Zheng, Hong and Hu, Yi and Chen, Xi and He, Liang},
  journal={arXiv preprint arXiv:2307.03952},
  year={2023}
}

@article{wang2024continuous,
  title={Continuous output personality detection models via mixed strategy training},
  author={Wang, Rong and Sun, Kun},
  journal={arXiv preprint arXiv:2406.16223},
  year={2024}
}

@inproceedings{li-etal-2025-big5,
    title = "{BIG}5-{CHAT}: Shaping {LLM} Personalities Through Training on Human-Grounded Data",
    author = "Li, Wenkai  and
      Liu, Jiarui  and
      Liu, Andy  and
      Zhou, Xuhui  and
      Diab, Mona T.  and
      Sap, Maarten",
    editor = "Che, Wanxiang  and
      Nabende, Joyce  and
      Shutova, Ekaterina  and
      Pilehvar, Mohammad Taher",
    booktitle = "Proceedings of the 63rd Annual Meeting of the Association for Computational Linguistics (Volume 1: Long Papers)",
    month = jul,
    year = "2025",
    address = "Vienna, Austria",
    publisher = "Association for Computational Linguistics",
    url = "https://aclanthology.org/2025.acl-long.999/",
    doi = "10.18653/v1/2025.acl-long.999",
    pages = "20434--20471",
    ISBN = "979-8-89176-251-0"
}

@inproceedings{li-etal-2024-evaluating-psychological,
    title = "Evaluating Psychological Safety of Large Language Models",
    author = "Li, Xingxuan  and
      Li, Yutong  and
      Qiu, Lin  and
      Joty, Shafiq  and
      Bing, Lidong",
    editor = "Al-Onaizan, Yaser  and
      Bansal, Mohit  and
      Chen, Yun-Nung",
    booktitle = "Proceedings of the 2024 Conference on Empirical Methods in Natural Language Processing",
    month = nov,
    year = "2024",
    address = "Miami, Florida, USA",
    publisher = "Association for Computational Linguistics",
    url = "https://aclanthology.org/2024.emnlp-main.108/",
    doi = "10.18653/v1/2024.emnlp-main.108",
    pages = "1826--1843"
}

@article{lu2023illuminating,
  title={Illuminating the black box: A psychometric investigation into the multifaceted nature of large language models},
  author={Lu, Yang and Yu, Jordan and Huang, Shou-Hsuan Stephen},
  journal={arXiv preprint arXiv:2312.14202},
  year={2023}
}

@article{tu4965442using,
  title={Using Large Language Models to Identify Narcissism Based on Texts},
  author={Tu, Zhuoran and Zhang, Zhaohui and Zhang, Wen and Luo, Fang and Bian, Ran},
  journal={Available at SSRN 4965442},
  year={2024}
}

@article{hadar2023plasticity,
  title={The plasticity of ChatGPT's mentalizing abilities: personalization for personality structures},
  author={Hadar-Shoval, Dorit and Elyoseph, Zohar and Lvovsky, Maya},
  journal={Frontiers in Psychiatry},
  volume={14},
  pages={1234397},
  year={2023},
  publisher={Frontiers Media SA}
}

@inproceedings{lee-etal-2025-llms,
    title = "Do {LLM}s Have Distinct and Consistent Personality? {TRAIT}: Personality Testset designed for {LLM}s with Psychometrics",
    author = "Lee, Seungbeen  and
      Lim, Seungwon  and
      Han, Seungju  and
      Oh, Giyeong  and
      Chae, Hyungjoo  and
      Chung, Jiwan  and
      Kim, Minju  and
      Kwak, Beong-woo  and
      Lee, Yeonsoo  and
      Lee, Dongha  and
      Yeo, Jinyoung  and
      Yu, Youngjae",
    editor = "Chiruzzo, Luis  and
      Ritter, Alan  and
      Wang, Lu",
    booktitle = "Findings of the Association for Computational Linguistics: NAACL 2025",
    month = apr,
    year = "2025",
    address = "Albuquerque, New Mexico",
    publisher = "Association for Computational Linguistics",
    url = "https://aclanthology.org/2025.findings-naacl.469/",
    doi = "10.18653/v1/2025.findings-naacl.469",
    pages = "8397--8437",
    ISBN = "979-8-89176-195-7"
}

@incollection{goldberg2013alternative,
  title={An alternative ``description of personality'': The Big-Five factor structure},
  author={Goldberg, Lewis R},
  booktitle={Personality and personality disorders},
  pages={34--47},
  year={2013},
  publisher={Routledge}
}

@book{myers1962myers,
  title={The myers-briggs type indicator},
  author={Myers, Isabel Briggs and others},
  volume={34},
  year={1962},
  publisher={Consulting Psychologists Press Palo Alto, CA}
}

@article{serapio2023personality,
  title={Personality Traits in Large Language Models},
  author={Serapio-Garc{\'\i}a, Greg and Safdari, Mustafa and Crepy, Cl{\'e}ment and Sun, Luning and Fitz, Stephen and Romero, Peter and Abdulhai, Marwa and Faust, Aleksandra and Matari{\'c}, Maja},
  journal={arXiv preprint arXiv:2307.00184},
  year={2023}
}

@article{shum2025big,
  title={Big Five Personality Trait Prediction Based on User Comments},
  author={Shum, Kit-May and Ptaszynski, Michal and Masui, Fumito},
  journal={Information},
  volume={16},
  number={5},
  pages={418},
  year={2025},
  publisher={MDPI}
}

@article{eysenck1991dimensions,
  title={Dimensions of personality: 16, 5 or 3?---Criteria for a taxonomic paradigm},
  author={Eysenck, Hans Jurgen},
  journal={Personality and individual differences},
  volume={12},
  number={8},
  pages={773--790},
  year={1991},
  publisher={Elsevier}
}

@article{paulhus2002dark,
  title={The dark triad of personality: Narcissism, Machiavellianism, and psychopathy},
  author={Paulhus, Delroy L and Williams, Kevin M},
  journal={Journal of research in personality},
  volume={36},
  number={6},
  pages={556--563},
  year={2002},
  publisher={Elsevier}
}

@article{john1991big,
  title={Big five inventory},
  author={John, Oliver P and Donahue, Eileen M and Kentle, Robert L},
  journal={Journal of personality and social psychology},
  year={1991}
}

@article{ashton2009hexaco,
  title={The HEXACO--60: A short measure of the major dimensions of personality},
  author={Ashton, Michael C and Lee, Kibeom},
  journal={Journal of personality assessment},
  volume={91},
  number={4},
  pages={340--345},
  year={2009},
  publisher={Taylor \& Francis}
}

@inproceedings{suh2024rediscovering,
  title={Rediscovering the Latent Dimensions of Personality with Large Language Models as Trait Descriptors},
  author={Suh, Joseph and Moon, Suhong and Kang, Minwoo and Chan, David},
  booktitle={NeurIPS 2024 Workshop on Behavioral Machine Learning},
  year={2024}
}

@article{li2025traits,
  title={Traits Run Deep: Enhancing Personality Assessment via Psychology-Guided LLM Representations and Multimodal Apparent Behaviors},
  author={Li, Jia and He, Yichao and Xu, Jiacheng and Luo, Tianhao and Hu, Zhenzhen and Hong, Richang and Wang, Meng},
  journal={arXiv preprint arXiv:2507.22367},
  year={2025}
}

@article{song2024identifying,
  title={Identifying multiple personalities in large language models with external evaluation},
  author={Song, Xiaoyang and Adachi, Yuta and Feng, Jessie and Lin, Mouwei and Yu, Linhao and Li, Frank and Gupta, Akshat and Anumanchipalli, Gopala and Kaur, Simerjot},
  journal={arXiv preprint arXiv:2402.14805},
  year={2024}
}

@article{huang2023revisiting,
  title={Revisiting the reliability of psychological scales on large language models},
  author={Huang, Jen-tse and Jiao, Wenxiang and Lam, Man Ho and Li, Eric John and Wang, Wenxuan and Lyu, Michael R},
  journal={arXiv preprint arXiv:2305.19926},
  year={2023}
}

@article{wen2024evaluating,
  title={Evaluating implicit bias in large language models by attacking from a psychometric perspective},
  author={Wen, Yuchen and Bi, Keping and Chen, Wei and Guo, Jiafeng and Cheng, Xueqi},
  journal={arXiv preprint arXiv:2406.14023},
  year={2024}
}

@article{zhao2025explicit,
  title={Explicit vs. implicit: Investigating social bias in large language models through self-reflection},
  author={Zhao, Yachao and Wang, Bo and Wang, Yan and Zhao, Dongming and He, Ruifang and Hou, Yuexian},
  journal={arXiv preprint arXiv:2501.02295},
  year={2025}
}

@article{ye2025generative,
  title={Generative psycho-lexical approach for constructing value systems in large language models},
  author={Ye, Haoran and Zhang, Tianze and Xie, Yuhang and Zhang, Liyuan and Ren, Yuanyi and Zhang, Xin and Song, Guojie},
  journal={arXiv preprint arXiv:2502.02444},
  year={2025}
}

@article{biedma2024beyond,
  title={Beyond human norms: Unveiling unique values of large language models through interdisciplinary approaches},
  author={Biedma, Pablo and Yi, Xiaoyuan and Huang, Linus and Sun, Maosong and Xie, Xing},
  journal={arXiv preprint arXiv:2404.12744},
  year={2024}
}

@INPROCEEDINGS{personality2021bert,
  author={Jun, He and Peng, Liu and Changhui, Jiang and Pengzheng, Liu and Shenke, Wu and Kejia, Zhong},
  booktitle={2021 IEEE International Conference on Emergency Science and Information Technology (ICESIT)}, 
  title={Personality Classification Based on Bert Model}, 
  year={2021},
  volume={},
  number={},
  pages={150-152},
  doi={10.1109/ICESIT53460.2021.9697048}}

@INPROCEEDINGS{awsep2022,
  author={Elourajini, Fahed and Aïmeur, Esma},
  booktitle={2022 IEEE International Conference on Data Mining Workshops (ICDMW)}, 
  title={AWS-EP: A Multi-Task Prediction Approach for MBTI/Big5 Personality Tests}, 
  year={2022},
  volume={},
  number={},
  pages={1-8},
  doi={10.1109/ICDMW58026.2022.00049}}

@inproceedings{huang2024reliability,
  title={On the reliability of psychological scales on large language models},
  author={Huang, Jen-tse and Jiao, Wenxiang and Lam, Man Ho and Li, Eric John and Wang, Wenxuan and Lyu, Michael},
  booktitle={Proceedings of the 2024 Conference on Empirical Methods in Natural Language Processing},
  pages={6152--6173},
  year={2024}
}

@article{cao2024worst,
  title={On the worst prompt performance of large language models},
  author={Cao, Bowen and Cai, Deng and Zhang, Zhisong and Zou, Yuexian and Lam, Wai},
  journal={Advances in Neural Information Processing Systems},
  volume={37},
  pages={69022--69042},
  year={2024}
}

@article{salinas2024butterfly,
  title={The butterfly effect of altering prompts: How small changes and jailbreaks affect large language model performance},
  author={Salinas, Abel and Morstatter, Fred},
  journal={arXiv preprint arXiv:2401.03729},
  year={2024}
}

@article{bodrovza2024personality,
  title={Personality testing of large language models: limited temporal stability, but highlighted prosociality},
  author={Bodro{\v{z}}a, Bojana and Dini{\'c}, Bojana M and Boji{\'c}, Ljubi{\v{s}}a},
  journal={Royal Society Open Science},
  volume={11},
  number={10},
  pages={240180},
  year={2024},
  publisher={The Royal Society}
}

@article{salecha2024large,
  title={Large language models display human-like social desirability biases in Big Five personality surveys},
  author={Salecha, Aadesh and Ireland, Molly E and Subrahmanya, Shashanka and Sedoc, Jo{\~a}o and Ungar, Lyle H and Eichstaedt, Johannes C},
  journal={PNAS nexus},
  volume={3},
  number={12},
  pages={pgae533},
  year={2024},
  publisher={Oxford University Press US}
}

@article{wang2025mbti,
  title={MBTI Personality Recognition and Performance Improvement in LLMs},
  author={Wang, EricRuiheng and Wang, Huandong},
  journal={Available at SSRN 5111274},
  year={2025}
}

@article{butcher2010minnesota,
  title={Minnesota multiphasic personality inventory},
  author={Butcher, James N},
  journal={The corsini encyclopedia of psychology},
  pages={1--3},
  year={2010},
  publisher={Wiley Online Library}
}

@inproceedings{ren2024valuebench,
  title={ValueBench: Towards Comprehensively Evaluating Value Orientations and Understanding of Large Language Models},
  author={Ren, Yuanyi and Ye, Haoran and Fang, Hanjun and Zhang, Xin and Song, Guojie},
  booktitle={Proceedings of the 62nd Annual Meeting of the Association for Computational Linguistics (Volume 1: Long Papers)},
  pages={2015--2040},
  year={2024}
}

@article{hogan1992hogan,
  title={Hogan Personality Inventory.},
  author={Hogan, Robert},
  journal={Psychological Test Bulletin},
  year={1992},
  publisher={Australian Council for Educational Research}
}

@article{gough1956california,
  title={California Psychological Inventory.},
  author={Gough, Harrison G},
  year={1956},
  publisher={Consulting Psychologists Press}
}

@article{chittem2025sac,
  title={SAC: A Framework for Measuring and Inducing Personality Traits in LLMs with Dynamic Intensity Control},
  author={Chittem, Adithya and Shrivastava, Aishna and Pendela, Sai Tarun and Challa, Jagat Sesh and Kumar, Dhruv},
  journal={arXiv preprint arXiv:2506.20993},
  year={2025}
}

@article{cattell2008sixteen,
  title={The Sixteen Personality Factor},
  author={Cattell, Heather EP and Mead, Alan D},
  journal={The SAGE Handbook of Personality Theory and Assessment: Personality Measurement and Testing (Volume 2)},
  volume={2},
  pages={135},
  year={2008},
  publisher={Sage}
}

@inproceedings{li2025can,
  title={Can large language models understand you better? an mbti personality detection dataset aligned with population traits},
  author={Li, Bohan and Guan, Jiannan and Dou, Longxu and Feng, Yunlong and Wang, Dingzirui and Xu, Yang and Wang, Enbo and Chen, Qiguang and Wang, Bichen and Xu, Xiao and others},
  booktitle={Proceedings of the 31st International Conference on Computational Linguistics},
  pages={5071--5081},
  year={2025}
}

@inproceedings{hartley2025personality,
  title={How personality traits shape llm risk-taking behaviour},
  author={Hartley, John and Hamill, Conor Brian and Seddon, Dale and Batra, Devesh and Okhrati, Ramin and Khraishi, Raad},
  booktitle={Findings of the Association for Computational Linguistics: ACL 2025},
  pages={21068--21092},
  year={2025}
}

@article{cohen2025exploring,
  title={Exploring Big Five Personality and AI Capability Effects in LLM-Simulated Negotiation Dialogues},
  author={Cohen, Myke C and Su, Zhe and Kao, Hsien-Te and Nguyen, Daniel and Lynch, Spencer and Sap, Maarten and Volkova, Svitlana},
  journal={arXiv preprint arXiv:2506.15928},
  year={2025}
}

@article{ashery2025emergent,
  title={Emergent social conventions and collective bias in LLM populations},
  author={Ashery, Ariel Flint and Aiello, Luca Maria and Baronchelli, Andrea},
  journal={Science Advances},
  volume={11},
  number={20},
  pages={eadu9368},
  year={2025},
  doi={10.1126/sciadv.adu9368}
}

@article{yan2024predicting,
  title={Predicting the big five personality traits in chinese counselling dialogues using large language models},
  author={Yan, Yang and Ma, Lizhi and Li, Anqi and Ma, Jingsong and Lan, Zhenzhong},
  journal={arXiv preprint arXiv:2406.17287},
  year={2024}
}

@article{zheng2025lmlpa,
  title={Lmlpa: Language model linguistic personality assessment},
  author={Zheng, Jingyao and Wang, Xian and Hosio, Simo and Xu, Xiaoxian and Lee, Lik-Hang},
  journal={Computational Linguistics},
  pages={1--42},
  year={2025},
  publisher={Mit press 255 Main Street, 9th Floor, Cambridge, Massachusetts 02142, USA~…}
}

@article{wang2025exploring,
  title={Exploring the impact of personality traits on llm bias and toxicity},
  author={Wang, Shuo and Li, Renhao and Chen, Xi and Yuan, Yulin and Wong, Derek F and Yang, Min},
  journal={arXiv preprint arXiv:2502.12566},
  year={2025}
}

@article{barua2024psychology,
  title={On the Psychology of GPT-4: Moderately anxious, slightly masculine, honest, and humble},
  author={Barua, Adrita and Brase, Gary and Dong, Ke and Hitzler, Pascal and Vasserman, Eugene},
  journal={arXiv preprint arXiv:2402.01777},
  year={2024}
}

@article{wang2025raters,
  title={Evaluating Large Language Models as Raters in Large-Scale Writing Assessments: A Psychometric Framework for Reliability and Validity},
  author={Wang, Yuehan and Huang, Jinyan and Du, Lun and Guo, Yuxin and Liu, Ying and Wang, Rong},
  journal={Computers and Education: Artificial Intelligence},
  pages={100481},
  year={2025},
  publisher={Elsevier}
}

@inproceedings{li2024evaluating,
  title={Evaluating Psychological Safety of Large Language Models},
  author={Li, Xingxuan and Li, Yutong and Qiu, Lin and Joty, Shafiq and Bing, Lidong},
  booktitle={Proceedings of the 2024 Conference on Empirical Methods in Natural Language Processing},
  pages={1826--1843},
  year={2024}
}

@article{laranjo2018conversational,
  title={Conversational agents in healthcare: a systematic review},
  author={Laranjo, Liliana and Dunn, Adam G and Tong, Huong Ly and Kocaballi, Ahmet Baki and Chen, Jessica and Bashir, Rabia and Surian, Didi and Gallego, Blanca and Magrabi, Farah and Lau, Annie YS and others},
  journal={Journal of the American Medical Informatics Association},
  volume={25},
  number={9},
  pages={1248--1258},
  year={2018},
  publisher={Oxford University Press}
}

@article{kim2023propile,
  title={Propile: Probing privacy leakage in large language models},
  author={Kim, Siwon and Yun, Sangdoo and Lee, Hwaran and Gubri, Martin and Yoon, Sungroh and Oh, Seong Joon},
  journal={Advances in Neural Information Processing Systems},
  volume={36},
  pages={20750--20762},
  year={2023}
}

@inproceedings{hong2024dp,
  title={DP-OPT: MAKE LARGE LANGUAGE MODEL YOUR PRIVACY-PRESERVING PROMPT ENGINEER},
  author={Hong, Junyuan and Wang, Jiachen T and Zhang, Chenhui and Li, Zhangheng and Li, Bo and Wang, Zhangyang},
  booktitle={12th International Conference on Learning Representations, ICLR 2024},
  year={2024}
}

@inproceedings{xiao2024large,
  title={Large language models can be contextual privacy protection learners},
  author={Xiao, Yijia and Jin, Yiqiao and Bai, Yushi and Wu, Yue and Yang, Xianjun and Luo, Xiao and Yu, Wenchao and Zhao, Xujiang and Liu, Yanchi and Gu, Quanquan and others},
  booktitle={Proceedings of the 2024 Conference on Empirical Methods in Natural Language Processing},
  pages={14179--14201},
  year={2024}
}

@inproceedings{santurkar2023opinions,
  title={Whose Opinions Do Language Models Reflect?},
  author={Santurkar, Shibani and Durmus, Esin and Ladd, Faisal and Han, Cinoo and Hashimoto, Tatsunori and Jurafsky, Dan},
  booktitle={Proceedings of the 40th International Conference on Machine Learning},
  pages={29971--30004},
  year={2023}
}

@article{zack2024assessing,
  title={Assessing the potential of {GPT-4} to perpetuate racial and gender biases in health care: a model evaluation study},
  author={Zack, Travis and Lehman, Eric and Suzgun, Mirac and Rodriguez, Jorge A and Celi, Leo Anthony and Gichoya, Judy and Jurafsky, Dan and Szolovits, Peter and Bates, David W and Abdulnour, Raja-Elie E and others},
  journal={The Lancet Digital Health},
  volume={6},
  number={1},
  pages={e12--e22},
  year={2024},
  publisher={Elsevier}
}

@article{navigli2023biases,
  title={Biases in Large Language Models: Origins, Inventory, and Discussion},
  author={Navigli, Roberto and Conia, Simone and Ross, Bj{\"o}rn},
  journal={Journal of Data and Information Quality},
  volume={15},
  number={2},
  pages={1--21},
  year={2023},
  publisher={ACM}
}

@inproceedings{kotek2023gender,
  title={Gender Bias and Stereotypes in Large Language Models},
  author={Kotek, Hadas and Dockum, Rikker and Sun, David Q},
  booktitle={Proceedings of the ACM Collective Intelligence Conference},
  pages={12--24},
  year={2023}
}

@article{muehlematter2021approval,
  title={Approval of artificial intelligence and machine learning-based medical devices in the {USA} and {Europe} (2015--20): a comparative analysis},
  author={Muehlematter, Urs J and Daniore, Paola and Vokinger, Kerstin N},
  journal={The Lancet Digital Health},
  volume={3},
  number={3},
  pages={e195--e203},
  year={2021},
  publisher={Elsevier}
}

@article{APA2013Telepsychology,
  title={Guidelines for the practice of telepsychology},
  author={{American Psychological Association}},
  journal={American Psychologist},
  volume={68},
  number={9},
  pages={791--800},
  year={2013},
  publisher={American Psychological Association}
}

@inproceedings{riemer2025tom_broken,
  title={Position: Theory of Mind Benchmarks are Broken for Large Language Models},
  author={Riemer, Matthew and Ashktorab, Zahra and Bouneffouf, Djallel and Das, Payel and Liu, Miao and Weisz, Justin D. and Campbell, Murray},
  booktitle={Proceedings of the 42nd International Conference on Machine Learning},
  series={Proceedings of Machine Learning Research},
  volume={267},
  pages={82091--82130},
  year={2025},
  publisher={PMLR},
  url={https://proceedings.mlr.press/v267/riemer25a.html}
}

@article{serapio2025psychometric,
  title={A psychometric framework for evaluating and shaping personality traits in large language models},
  author={Serapio-Garc{\'\i}a, Greg and Safdari, Mustafa and Crepy, Cl{\'e}ment and Fitz, Stephen and Romero, Peter and Sun, Luning and Abdulhai, Marwa and Faust, Aleksandra and Matari{\'c}, Maja},
  journal={Nature Machine Intelligence},
  volume={7},
  pages={329--343},
  year={2025},
  publisher={Nature Publishing Group}
}

@article{badawi2025trust_mental,
  title={When Can We Trust {LLMs} in Mental Health? Large-Scale Benchmarks for Reliable {LLM} Evaluation},
  author={Badawi, Abeer and Rahimi, Elahe and Laskar, Md Tahmid Rahman and Grach, Sheri and Bertrand, Lindsay and Danok, Lames and Huang, Jimmy and Rudzicz, Frank and Dolatabadi, Elham},
  journal={arXiv preprint arXiv:2510.19032},
  year={2025}
}

@article{shen2024cultural_bias,
  title={Cultural bias and cultural alignment of large language models},
  author={Shen, Dinghan and Liscio, Enrico and Murukannaiah, Pradeep K and others},
  journal={PNAS Nexus},
  volume={3},
  number={9},
  pages={pgae346},
  year={2024},
  publisher={Oxford University Press}
}

@inproceedings{chen2025tom_survey_acl,
  title={Theory of Mind in Large Language Models: Assessment and Enhancement},
  author={Chen, Ruirui and Jiang, Weifeng and Qin, Chengwei and Tan, Cheston},
  booktitle={Proceedings of the 63rd Annual Meeting of the Association for Computational Linguistics (Volume 1: Long Papers)},
  pages={31539--31558},
  year={2025},
  address={Vienna, Austria},
  publisher={Association for Computational Linguistics},
  doi={10.18653/v1/2025.acl-long.1522},
  url={https://aclanthology.org/2025.acl-long.1522/}
}

\begin{IEEEbiographynophoto}{Yudong Li}
received the Ph.D. degree from Shenzhen University, Shenzhen, China. He is currently a Research Associate with Tsinghua University, Beijing, China. His research interests include pre-trained language models, multimodal pre-training, and applications of large language models.
\end{IEEEbiographynophoto}

\begin{IEEEbiographynophoto}{Xiaoyi Chen}
received the bachelor's degree from Shenzhen University, Shenzhen, China. She is currently pursuing the master's degree in computer science.
\end{IEEEbiographynophoto}

\begin{IEEEbiographynophoto}{Jiawei Cai}
received the bachelor's degree from Shenzhen University, Shenzhen, China. He is currently pursuing the master's degree in computer science.
\end{IEEEbiographynophoto}

\begin{IEEEbiographynophoto}{Zehao Zhong}
received the bachelor's degree from Shenzhen University, Shenzhen, China. He is currently pursuing the master's degree in computer science.
\end{IEEEbiographynophoto}

\begin{IEEEbiographynophoto}{Haoyang Yang}
received the bachelor's degree from Shenzhen University, Shenzhen, China. He is currently pursuing the master's degree in computer science.
\end{IEEEbiographynophoto}

\begin{IEEEbiographynophoto}{Huajin Tang}
(Fellow, IEEE) received the B.Eng. degree from Zhejiang University, Hangzhou, China in 1998, the M.Eng. degree from Shanghai Jiao Tong University, Shanghai, China in 2001, and the Ph.D. degree from the National University of Singapore, Singapore in 2005. He was a System Engineer with STMicroelectronics, Singapore, from 2004 to 2006. From 2006 to 2008, he was a Post-Doctoral Fellow with Queensland Brain Institute, The University of Queensland, Brisbane, QLD, Australia. Since 2008, he has been the Head of the Robotic Cognition Laboratory, Institute for Infocomm Research, Agency for Science, Technology and Research (A*STAR), Singapore. Since 2014, he has been a Professor with the College of Computer Science, Sichuan University, China. He is currently a Professor with the College of Computer Science and Technology, Zhejiang University. His current research interests include neuromorphic computing, neuromorphic hardware and cognitive systems, and robotic cognition. Dr. Tang is a Board of Governors Member of the International Neural Networks Society. He received the 2016 IEEE Outstanding TNNLS Paper Award, the 2019 IEEE Computational Intelligence Magazine Outstanding Paper Award, and 2023 Neural Networks Best Paper Award. He has served as an Associate Editor for IEEE TRANSACTIONS ON NEURAL NETWORKS AND LEARNING SYSTEMS, IEEE TRANSACTIONS ON COGNITIVE AND DEVELOPMENTAL SYSTEMS, Frontiers in Neuromorphic Engineering, and Neural Networks.
\end{IEEEbiographynophoto}

\begin{IEEEbiographynophoto}{Linlin Shen}
(Senior Member, IEEE) is currently a Pengcheng Scholar Distinguished Professor at School of Artificial Intelligence, Shenzhen University, Shenzhen, China. He is also a Honorary professor at School of Computer Science, University of Nottingham, UK. He serves as the Deputy director of National Engineering Lab of Big Data Computing Technology, Director of Computer Vision Institute, AI Research Center for Medical Image Analysis and Diagnosis and China-UK joint research lab for visual information processing. He also serves as the Co-Editor-in-Chief of the IET journal of Cognitive Computation and Systems, Senoir Area Editor of IEEE Trans. on Image Processing, Senior Editor of Expert Systems With Applications, and Associate Editor of Pattern Recognition and Scieitific Data. His research interests include deep learning, facial recognition, analysis/synthesis and medical image processing. Prof. Shen is listed as the“Most Cited Chinese Researchers”by Elsevier, “Top 0.05\% Highly Ranked Scholar”by ScholarGPS, and listed in a ranking of the“Top 2\% Scientists in the World” by Stanford University. He received the“Best Paper Runner-up Award”from the journal of IEEE Transactions on Affective Computing, "Top Cited Article" from Wiley, and “Most Cited Paper Award” from the journal of Image and Vision Computing. His cell classification algorithms were the winners of the International Contest on Pattern Recognition Techniques for Indirect Immunofluorescence Images held by ICIP and ICPR. His team has also been the runner-up and second runner-up of a number of competitions for object detection in remote sensing images, nucleus detection in histopathology images and facial expression recognition.
\end{IEEEbiographynophoto}

\end{document}